\documentclass[a4paper,12pt]{article}
\usepackage{amsmath}
 \pdfoutput=1
\usepackage{enumitem}
\usepackage{color,pxfonts,fix-cm}
\usepackage{latexsym}
\usepackage[absolute,overlay]{textpos}
\usepackage[mathletters]{ucs}
\usepackage[T1]{fontenc}
\usepackage[utf8x]{inputenc}
\usepackage{pict2e}
\usepackage{wasysym}
\usepackage[english]{babel}
\usepackage{tikz}

\begin{document}

\begin{center}
     Low energy models of string theory\\
     \vspace{5mm}
    Poula Tadros and Iiro Vilja\\
\vspace{5mm}
Department of Physics and Astronomy, University of Turku, 20014, Turku,Finland.\\
\vspace{5mm}

Emails: poulatadros9@gmail.com , Vilja@utu.fi
\end{center}
\begin{abstract}
String theory is the prime candidate for the theory of everything. However, it must be defined in ten dimensions to be consistent. To get 4D physics, the 6 other dimensions should be curled up in a small compact manifold, this procedure is called string compactification. In this review, we will review different compactification schemes proving that in absence of flux, the compact manifold must be a Calabi-Yau manifold. Then, we review compactifications with flux using generalized complex geometry. We then discuss some applications in cosmology like the swampland project and the cosmological models derived from it. We then discuss non relativistic string theories and introduce a toroidal compactifications for such theories. Finally, we discuss some open questions in the field.
\end{abstract}

\newpage

\tableofcontents
\section{Introduction}
String theory is currently the most promising candidate for the theory of everything but it contains a conformal anomaly which means an inconsistency in the theory. To cancel this anomaly bosonic string theory must be defined on a 26 dimensional spacetime and superstring theories have to be defined on a 10 dimensional spacetime. But we know that we live in 4 dimensions so we have to reduce the dimensions of string theories spacetimes to 4. The way to do this is to assume that the ambient spacetime manifold of the string theory is a product of two manifolds: a 4 dimensional non compact manifold which will represent the 4 dimensional spacetime we observe, and a higher dimensional compact manifold (22 dimensional in the case of bosonic string theory and 6 dimensional in case of superstring theories) representing an internal space which we can not observe because the compact manifold is too small for us to observe in low energy, this procedure is called a string compactification.

String compactification models usually give rise to massless scalar fields with no potentials on the 4 dimensional manifolds, these scalars are called moduli and are problematic from the physical point of view because they can mediate long range forces which are not observed in nature. To solve this problem, additional structures are added to the ambient manifold namely differential forms called fluxes. Fluxes induce potentials to moduli so that they can not mediate long range forces, this process is called moduli stabilization. String compactifications on manifolds with fluxes are called flux compactifications.

On studying string compactification models on a certain manifold, there are three main steps:
\begin{enumerate}
    \item The study of the possible compactification manifolds and their geometry i.e. we study which manifold we can compactify the theory on to give a realistic 4 dimensional particle physics theory, and the spectrum of this low energy 4 dimensional model since the geometry of the internal manifold affects the 4 dimensional low energy model.
    \item The classification of vacua in the compactified model i.e. to study what are the possible vacua can arise from the theory and see which one makes a physically viable 4 dimensional model. This step is particularly important if we want to construct cosmological models from the compactification model as we will see in chapter 8.
    \item The study of moduli spaces of the compactification models. This is very important because the geometry of the moduli space dramatically affects the 4 dimensional model and it also gives hints of which flux should we introduce to stabilize the moduli.
\end{enumerate}
In this review we focus on step 1, and only get in touch with step 2 in chapters 6.

In chapter 2, we review some mathematical concepts that will be used throughout the review. In chapter 3, we introduce some string compactification models and study the low energy 4 dimensional models arising from them and their spectra, since there are many compactification models, we introduce the most generic ones so the reader can follow the same procedure with other similar models. We study the process for bosonic string theory as a toy model, type II and heterotic superstring theories. Then, we present some challenges to the aforementioned models leading to choosing the compactification manifold to be a Calabi-Yau. In chapter 4, we study compactifications on Calabi-Yau compactification i.e. string compactifications with the internal manifold to be Calabi-Yau. There we show that a realistic low energy model from string compactifications without flux must arise from a Calabi-Yau compactification in case of heterotic supertring theories, we also present the compactification schemes of bosonic strings, type II superstring theories and non supersymmetric string theory. However, as mentioned earlier, this does not stabilize the moduli in the theory thus, flux must be added. In chapter 5 we review generalised complex geometry which is the math used in flux compactifications. In chapter 6, we study different models of flux compactifications showing how flux can break supersymmetry leading to realistic low energy models. In chapter 7, we discuss the swampland project which was designed to test whether a low energy model is consistent with quantum gravity, in this case string theory, and we present some vacua from string compactification models motivated by the swampland conjecture representing possible realistic cosmological models. In chapter 8, we review non realtivistic string theory, so far there are no compactification models for non relativistic strings, we construct a toroidal compactification model in this chapter and directions for future research in the field. In chapter 9, we present some open problems and what are the directions of research at the moment.

\section{Mathematical tools}
\subsection{Basic category theory}
\subsubsection{Categories and functors}

Category theory was founded by Eilenberg and Maclane in 1945 
[1] as a step to define natural transformations between mathematical structures. Category theory can be though of as a unifying scheme that takes a bird's eye view on all mathematical structures, for example we can define sets and functions between them, groups and homomorphisms preserving group structure, topological spaces and continuous maps (preserving the topological structure), rings and ring homomorphisms, vector spaces and linear transformations, etc. The main idea is a structure and a map preserving it thus, we can unify them calling the structure "objects" and the maps "morphisms" and the whole system of objects and morphisms as Category.\\
There are two equivalent ways to properly define Categories. The first uses the notions of metagraphs and metacategories as presented in [2]. A recent way defines Categories directly by some axioms. In this review, we use the second approach.

\textbf{Definition:} A \textbf{Category} $\mathcal{C}$ consists of:
\begin{enumerate}
\item Objects denoted by capital letters A,B,... collectively denoted by obj($\mathcal{C}$).
\item morphisms between objects denoted by small letters f,g,... . Morphisms between A and B in the category $\mathcal{C}$ are denoted by $\mathcal{C}(A,B)$ or Hom(A,B).  A is called the domain and B the codomain. The collection of all morphism in the category $\mathcal{C}$ is denoted by Hom($\mathcal{C}$)
\end{enumerate}
Satisfying the following axioms:
\begin{enumerate}
\item Composition of morphisms is defined i.e. the map $ \mathcal{C}(A,B) \times \mathcal{C}(B,C) \rightarrow \mathcal{C}(A,C)$ ; $(f,g) \rightarrow f \circ g$ is defined.
\item Associativity of morphisms i.e. $\forall f \in \mathcal{C}(A,B) , g \in \mathcal{C}(B,C) $ and $h \in \mathcal{C}(C,D)$ ; $(f \circ g) \circ h = f \circ (g \circ h)$.
\item The existence of identity morphism to each object i.e. $ \forall A \in \mathcal{C}$ $ \exists \  1_A \in \mathcal{C}(A,A)$; $\forall f \in \mathcal{C}(A,B)$ $f \circ 1_A =f= 1_B \circ f$.
\end{enumerate}
A morphism f from A to B is called invertible if there exists another morphism g from B to A such that $f\circ g= 1_B$ and $g \circ f = 1_A$.

Graphically, we can consider objects as points and morphisms as arrows as shown in fig.1

\begin{figure}
    \centering

\tikzset{every picture/.style={line width=0.75pt}} 

\begin{tikzpicture}[x=0.75pt,y=0.75pt,yscale=-1,xscale=1]

\draw   (226.03,185.25) .. controls (226.03,138.87) and (279.81,101.27) .. (346.15,101.27) .. controls (412.5,101.27) and (466.28,138.87) .. (466.28,185.25) .. controls (466.28,231.64) and (412.5,269.24) .. (346.15,269.24) .. controls (279.81,269.24) and (226.03,231.64) .. (226.03,185.25) -- cycle ;
\draw    (299.24,165.46) -- (365.97,166.32) ;
\draw [shift={(367.97,166.34)}, rotate = 180.73] [color={rgb, 255:red, 0; green, 0; blue, 0 }  ][line width=0.75]    (10.93,-3.29) .. controls (6.95,-1.4) and (3.31,-0.3) .. (0,0) .. controls (3.31,0.3) and (6.95,1.4) .. (10.93,3.29)   ;
\draw    (291.59,163.43) .. controls (248.24,129.69) and (342.72,121.54) .. (300.56,164.15) ;
\draw [shift={(299.24,165.46)}, rotate = 315.84000000000003] [color={rgb, 255:red, 0; green, 0; blue, 0 }  ][line width=0.75]    (10.93,-3.29) .. controls (6.95,-1.4) and (3.31,-0.3) .. (0,0) .. controls (3.31,0.3) and (6.95,1.4) .. (10.93,3.29)   ;
\draw    (375.73,163.43) .. controls (341.76,113.52) and (429.73,122.14) .. (384.46,164.9) ;
\draw [shift={(383.05,166.2)}, rotate = 317.85] [color={rgb, 255:red, 0; green, 0; blue, 0 }  ][line width=0.75]    (10.93,-3.29) .. controls (6.95,-1.4) and (3.31,-0.3) .. (0,0) .. controls (3.31,0.3) and (6.95,1.4) .. (10.93,3.29)   ;
\draw    (335.49,229.25) .. controls (393.01,191.24) and (297.67,173.83) .. (325.47,223.16) ;
\draw [shift={(326.35,224.68)}, rotate = 239.12] [color={rgb, 255:red, 0; green, 0; blue, 0 }  ][line width=0.75]    (10.93,-3.29) .. controls (6.95,-1.4) and (3.31,-0.3) .. (0,0) .. controls (3.31,0.3) and (6.95,1.4) .. (10.93,3.29)   ;

\draw (306.96,276.93) node [anchor=north west][inner sep=0.75pt]   [align=left] {Category};
\draw (288.41,157.86) node [anchor=north west][inner sep=0.75pt]   [align=left] {A};
\draw (371.8,158.32) node [anchor=north west][inner sep=0.75pt]   [align=left] {B};
\draw (322.84,220.91) node [anchor=north west][inner sep=0.75pt]   [align=left] {C};
\draw (325.04,149.68) node [anchor=north west][inner sep=0.75pt]   [align=left] {f};

\end{tikzpicture}
\caption{A graphical representation of a Category.}
    
    \label{figure 1}
\end{figure}
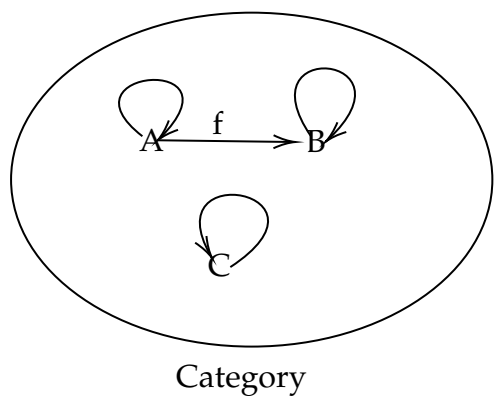

Examples : 
\begin{enumerate}
\item \textbf{Set}: The Category of sets (sets are objects and functions are morphisms).
\item \textbf{Grp}: The category of groups (groups are objects and homomorphisms are morphisms).
\item \textbf{Ab} : the category of Abelian groups.
\item \textbf{Top} : The category of topological spaces with  continuous maps as morphisms.
\item $\mathbf{Vect_k} $: The category of vector spaces over the field k with linear transformations as morphisms.
\end{enumerate}

\textbf{Definition:} A category is called \textbf{small} if the collection of objects and the collection of morphisms form sets. If a category is not small, it is called large.

\textbf{Example:} Any finite category is small.

\textbf{Non-Example:} The category of sets (\textbf{Set}) is a large category because the collection of sets is not a set.

\textbf{Definition}: A \textbf{Monoid} is a category with one object.

\textbf{Example:} A group is a monoid (the set is the object and group operation between elements are morphisms )such that every element has an inverse.

Next we consider mapping between categories.

\textbf{Definition:} A \textbf{Covariant functor} F from a category $\mathcal{C}$ to a category $\mathcal{A}$ is a mapping assigning objects in $\mathcal{C}$ to objects in $\mathcal{A}$, and morphisms in $\mathcal{C}$ to morphisms in $\mathcal{A}$, satisfying the following axioms:
\begin{enumerate}
\item $F(1_A)=1_{F(A)}$ for all A in $\mathcal{C}$.
\item $\forall f,g \in Hom(C) \ F(fg)=F(f)F(g)$ so called covariant.
\end{enumerate}

\textbf{Examples:}
\begin{enumerate}
\item The assignment of a fundamental group to topological spaces is considered as a covariant functor from \textbf{Top} to \textbf{Ab}
\item An important example for string theory is that moduli are defined as covariant functors from the category of schemes to the category of sets as defined later.
\end{enumerate}
\subsubsection{Diagrams}
Calculations using category theory uses diagrams modelling objects as points and morphisms as arrows. This method helps writing definitions and theorems in a more compact way which is easier to handle.

\textbf{Definition.}A diagram is called \textbf{commutative} if composition rules does not depend on a path in the diagram.

\textbf{Example:} The statement "Diagram 2 commutes" (assuming every arrow is invertible)is equivalent to the conditions: $f=g^{-1} \circ  I \circ h$ , $g \circ f = I \circ h$ , $h=I^{-1} \circ g \circ f$ , $I=g \circ f \circ h^{-1}$ , $g=I \circ h \circ f^{-1}$ and the inverse counterparts of these conditions.\\
This representation is the representation used throughout the review.

\begin{figure}
    \centering

\tikzset{every picture/.style={line width=0.75pt}} 

\begin{tikzpicture}[x=0.75pt,y=0.75pt,yscale=-1,xscale=1]

\draw    (104.67,59.18) -- (236.85,59.4) ;
\draw [shift={(238.85,59.4)}, rotate = 180.1] [color={rgb, 255:red, 0; green, 0; blue, 0 }  ][line width=0.75]    (10.93,-3.29) .. controls (6.95,-1.4) and (3.31,-0.3) .. (0,0) .. controls (3.31,0.3) and (6.95,1.4) .. (10.93,3.29)   ;
\draw    (99.13,73.46) -- (99.01,151.63) ;
\draw [shift={(99.01,153.63)}, rotate = 270.08] [color={rgb, 255:red, 0; green, 0; blue, 0 }  ][line width=0.75]    (10.93,-3.29) .. controls (6.95,-1.4) and (3.31,-0.3) .. (0,0) .. controls (3.31,0.3) and (6.95,1.4) .. (10.93,3.29)   ;
\draw    (250.01,74.46) -- (250.01,148.36) ;
\draw [shift={(250.01,150.36)}, rotate = 270] [color={rgb, 255:red, 0; green, 0; blue, 0 }  ][line width=0.75]    (10.93,-3.29) .. controls (6.95,-1.4) and (3.31,-0.3) .. (0,0) .. controls (3.31,0.3) and (6.95,1.4) .. (10.93,3.29)   ;
\draw    (111.12,166.34) -- (231.02,167.33) ;
\draw [shift={(233.02,167.34)}, rotate = 180.47] [color={rgb, 255:red, 0; green, 0; blue, 0 }  ][line width=0.75]    (10.93,-3.29) .. controls (6.95,-1.4) and (3.31,-0.3) .. (0,0) .. controls (3.31,0.3) and (6.95,1.4) .. (10.93,3.29)   ;

\draw (92.74,51.56) node [anchor=north west][inner sep=0.75pt]   [align=left] {A};
\draw (242.99,50.61) node [anchor=north west][inner sep=0.75pt]   [align=left] {B};
\draw (90.94,157.84) node [anchor=north west][inner sep=0.75pt]   [align=left] {C};
\draw (242.56,158.17) node [anchor=north west][inner sep=0.75pt]   [align=left] {D};
\draw (163.97,43.12) node [anchor=north west][inner sep=0.75pt]   [align=left] {f};
\draw (255.9,98.06) node [anchor=north west][inner sep=0.75pt]   [align=left] {g};
\draw (84.03,102.06) node [anchor=north west][inner sep=0.75pt]   [align=left] {h};
\draw (157.97,150) node [anchor=north west][inner sep=0.75pt]   [align=left] {I};

\end{tikzpicture}
  \caption{This is an example of a diagram in category theory, this id labeled in the text by "Diagram 2".}
    
\end{figure}
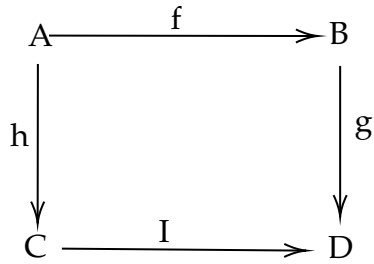

\subsubsection{Natural transformations}

Natural transformations are transformations between functors between two specified categories. It can be thought of as morphisms between functors motivating the definition of functor categories. We will not discuss functor categories in this review.

\textbf{Definition:} Let s and t be functors from the category $\mathcal{C}$ to the category $\mathcal{A}$, and let C be an arbitrary object in $\mathcal{C}$, and f be a morphism in $\mathcal{C}$ mapping C to C'. A \textbf{Natural transformation} $\tau :s \rightarrow t$ is a mapping assigning to each object C in $\mathcal{C}$ an arrow from s(C) to t(C) such that diagram 3 commutes.
\begin{figure}
    \centering

\tikzset{every picture/.style={line width=0.75pt}} 

\begin{tikzpicture}[x=0.75pt,y=0.75pt,yscale=-1,xscale=1]

\draw    (104.67,59.18) -- (236.85,59.4) ;
\draw [shift={(238.85,59.4)}, rotate = 180.1] [color={rgb, 255:red, 0; green, 0; blue, 0 }  ][line width=0.75]    (10.93,-3.29) .. controls (6.95,-1.4) and (3.31,-0.3) .. (0,0) .. controls (3.31,0.3) and (6.95,1.4) .. (10.93,3.29)   ;
\draw    (99.13,73.46) -- (99.01,151.63) ;
\draw [shift={(99.01,153.63)}, rotate = 270.08] [color={rgb, 255:red, 0; green, 0; blue, 0 }  ][line width=0.75]    (10.93,-3.29) .. controls (6.95,-1.4) and (3.31,-0.3) .. (0,0) .. controls (3.31,0.3) and (6.95,1.4) .. (10.93,3.29)   ;
\draw    (250.01,74.46) -- (250.01,148.36) ;
\draw [shift={(250.01,150.36)}, rotate = 270] [color={rgb, 255:red, 0; green, 0; blue, 0 }  ][line width=0.75]    (10.93,-3.29) .. controls (6.95,-1.4) and (3.31,-0.3) .. (0,0) .. controls (3.31,0.3) and (6.95,1.4) .. (10.93,3.29)   ;
\draw    (111.12,166.34) -- (231.02,167.33) ;
\draw [shift={(233.02,167.34)}, rotate = 180.47] [color={rgb, 255:red, 0; green, 0; blue, 0 }  ][line width=0.75]    (10.93,-3.29) .. controls (6.95,-1.4) and (3.31,-0.3) .. (0,0) .. controls (3.31,0.3) and (6.95,1.4) .. (10.93,3.29)   ;

\draw (74,48.92) node [anchor=north west][inner sep=0.75pt]   [align=left] {s(C)};
\draw (242.99,50.61) node [anchor=north west][inner sep=0.75pt]   [align=left] {t(C)};
\draw (75,156.52) node [anchor=north west][inner sep=0.75pt]   [align=left] {s(C')};
\draw (242.56,158.17) node [anchor=north west][inner sep=0.75pt]   [align=left] {t(C')};
\draw (160.67,41.14) node [anchor=north west][inner sep=0.75pt]   [align=left] {$\displaystyle \tau $c};
\draw (255.9,98.06) node [anchor=north west][inner sep=0.75pt]   [align=left] {t(f)};
\draw (71.39,99.42) node [anchor=north west][inner sep=0.75pt]   [align=left] {s(f)};
\draw (159.95,146.7) node [anchor=north west][inner sep=0.75pt]   [align=left] {$\displaystyle \tau c'$};

\end{tikzpicture}

    \caption{This Diagram represents the action of natural transformations if it commutes. It is refered to in the text by "Diagram 3".}
    
\end{figure}
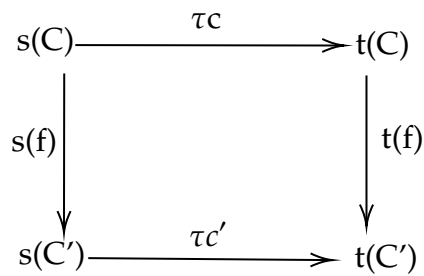

A natural transformation with every object $\tau$c is invertible is called a natural equivalence.

\textbf{Example:} The most famous example of a natural transformation is a determinant of matrices with entries in an arbitrary commutative ring, which is a natural transformation between two functors from the category of commutative rings to the category of groups.

\subsubsection{The duality principle}

The informal statement of the duality principle is that any result can be deduced from category theory has a dual statement where all arrows are reversed i.e. domains and codomains of all morphisms are interchanged.

\textbf{Definition:} For any category $\mathcal{C}$, we can define its opposite category denoted by $\mathcal{C}^{op}$ consisting of the same objects but all arrows are reversed. If $\mathcal{C}$ = $\mathcal{C}^{op}$ then it is called \textbf{self dual}.

\textbf{Definition:} A functor of the form G:$\mathcal{C}^{op} \rightarrow \mathcal{A}$ is called a \textbf{contravariant functor} from $\mathcal{C}$ to $\mathcal{A}$.

It is called contravariant because the composition rule in this case is $G(f \circ g)= G(g) \circ G(f)$, because all morphisms in $\mathcal{C}$ are reversed. Note that a functor from $\mathcal{C}^{op}$ to $\mathcal{A}^{op}$ is covariant because the composition rule is reversed two times.

\textbf{Example:} The assignment of cohomology groups to each topological space is a contravariant functor form $\mathbf{Top}$ to \textbf{Ab}.

\textbf{The formal duality principle}:
For any sentence in the language of category theory following from axioms of categories, there is a dual statement which also follows from the same axioms.

\subsubsection{Yoneda lemma}

Yoneda lemma is considered one of the most important results in category theory. It basically says that we can identify arbitrary objects in arbitrary categories knowing only their relations with other categories, for example, we can identify a topological space by knowing how the other topological spaces are mapped to it continuously.

\textbf{Definitions:}
\begin{enumerate}
\item Assume that $\mathcal{C}$ is a locally small category. We define the \textbf{$Hom_{\mathcal{C}}(-,C)$} as a contravariant functor from a category $\mathcal{C}$ to \textbf{Set}, sending an object d in $\mathcal{C}$ to the set Hom(D,C) the set of all morphisms from D to C, and sending morphisms in $\mathcal{C}$ to functions in \textbf{Set} according to the following rule : \\
If $f:D \rightarrow E$ is a morphism from D to E in $\mathcal{C}$, Then $Hom_{\mathcal{C}}(f,C)$ is a function from Hom(E,C) to Hom(D,C) as the composition of the morphism form E to C and f.

\item Let F be a covariant functor from $\mathcal{C}$ to $\mathcal{A}$ (locally small categories). F induces a function
$$F_{C,A}:Hom_{\mathcal{C}}(C,A)\rightarrow Hom_{\mathcal{A}(F(C),F(A))}$$
F is called \textbf{faithful} if $F_{C,A}$ is injective, and is called \textbf{full} if $F_{C,A}$ is surjective. If $F_{C,A}$ is bijective then F is called \textbf{Fully faithful}.

\item We denote the category of all functors from $\mathcal{C}$ to $\mathcal{A}$, with morphisms are natural transformations, by \textbf{Func($\mathcal{C}$,$\mathcal{D}$)}.

\item Let $\mathcal{C}$ be a locally small category. \textbf{Yoneda functor} is defined as $h:\mathcal{C} \rightarrow \mathbf{Func(\mathcal{C}^{op},Set)}$, mapping objects as $C \rightarrow Hom(- , C)$ and morphisms as a covariant version of definition 1(note that Yoneda functor is covariant).
\end{enumerate}

\textbf{Lemma(Yoneda)}:
Let $\mathcal{C}$ be a locally small category. For every object 
C in $\mathcal{C}$ and for every functor F in \textbf{Func($\mathcal{C}^{op}$, \textbf{Set})}, there is an isomorphism 
$$Hom(hC,F)\cong FC,$$
where h is the Yoneda functor. Moreover, the isomorphism is natural in the sense discussed before.

\textbf{Theorem}:
The Yoneda functor is fully faithful.

Yoneda's lemma tells us that all information of an object in a category, say an unknown manifold we want to study, is entirely embedded in the set of functors with other objects, in our example if we know how to map other manifolds to this unknown manifold, then we can determine its structure entirely.

\subsection{Manifolds}
Here, we assume the reader to be familiar with the definition of real and complex manifolds, tangent and cotangent spaces, tensors and lie groups. This section follows the analysis of Ref [3,4].
\subsubsection{De Rham cohomology}

\textbf{Definition:}A differential $\omega$ form is called exact if there exists another differential form $\Omega$ such that $d\Omega=\omega$, and is called closed if $d\omega=0$.

It is easy to prove that the vector space of exact forms is a subspace of the vector space of closed forms. Their quotient is called \textbf{De Rham cohomology}, denoted by $H^k(M)$ (k is the degree of the forms and M is the manifold).

\textbf{Proposition:} If the dimension of the manifold is n then $k>n \implies H^k(M)=0$.

An important property of De Rham cohomology is that we can define $H^*(M)=\bigoplus _{k=0}^{n}H^k(M)$ which can be shown to be an anticommutative graded ring over R. Thus, De Rham cohomology is a contravariant functor from the category of $C^{\infty}$ manifolds to the category of anticommutative graded rings.

\subsubsection{Vector bundles}

\textbf{Definition:} A smooth \textbf{k-dimensional vector bundle} is a pair of smooth manifolds E (called the total space), and M (called the base) with a surjective map $\pi : E \rightarrow M$ (called the projection), such that
\begin{enumerate}

\item $\forall p \in M \ \ E_p=\pi^{-1}(p)$ (called a fiber of E over p) is a vector space.

\item $\forall p \in M \ \  \exists$ a neighborhood U of p such that U and the product $\text U \times \mathbb{R}^K$ are diffeomorphic i.e. there exists a mapping $\phi : \pi^{-1}(\text U) \rightarrow \text U \times \mathbb{R} ^k$ such that $\phi$ is a diffeomorphism and diagram 4 commutes. In this case $\phi$ called a local trivialization of E, 

\begin{figure}
\centering

\tikzset{every picture/.style={line width=0.75pt}} 

\begin{tikzpicture}[x=0.75pt,y=0.75pt,yscale=-1,xscale=1]

\draw    (124.11,91.44) -- (195.76,176.8) ;
\draw [shift={(197.05,178.33)}, rotate = 229.99] [color={rgb, 255:red, 0; green, 0; blue, 0 }  ][line width=0.75]    (10.93,-3.29) .. controls (6.95,-1.4) and (3.31,-0.3) .. (0,0) .. controls (3.31,0.3) and (6.95,1.4) .. (10.93,3.29)   ;
\draw    (156.08,74.46) -- (270.99,74.46) ;
\draw [shift={(272.99,74.46)}, rotate = 180] [color={rgb, 255:red, 0; green, 0; blue, 0 }  ][line width=0.75]    (10.93,-3.29) .. controls (6.95,-1.4) and (3.31,-0.3) .. (0,0) .. controls (3.31,0.3) and (6.95,1.4) .. (10.93,3.29)   ;
\draw    (298.97,92.44) -- (225.31,180.79) ;
\draw [shift={(224.03,182.32)}, rotate = 309.82] [color={rgb, 255:red, 0; green, 0; blue, 0 }  ][line width=0.75]    (10.93,-3.29) .. controls (6.95,-1.4) and (3.31,-0.3) .. (0,0) .. controls (3.31,0.3) and (6.95,1.4) .. (10.93,3.29)   ;

\draw (101.64,64.1) node [anchor=north west][inner sep=0.75pt]   [align=left] {$\displaystyle \pi ^{-1}$(U)};
\draw (205.05,178.33) node [anchor=north west][inner sep=0.75pt]   [align=left] {U$ $};
\draw (279.77,64.08) node [anchor=north west][inner sep=0.75pt]   [align=left] {U $\displaystyle \times \ R^{k}$};
\draw (267.89,134.42) node [anchor=north west][inner sep=0.75pt]    {$\pi _{1}$};
\draw (203.94,56.1) node [anchor=north west][inner sep=0.75pt]   [align=left] {$\displaystyle \varphi $};
\draw (137.99,130.02) node [anchor=north west][inner sep=0.75pt]   [align=left] {$\displaystyle \pi $};

\end{tikzpicture}
\caption{This diagram's commutativity represents the condition of local trivialization existence and it is refered to in the text by "Diagram 4".}

\end{figure}
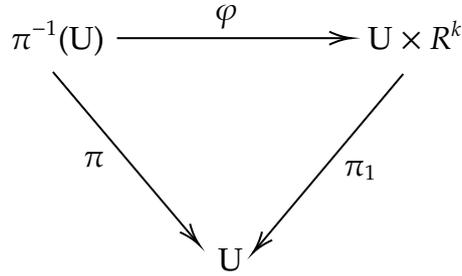
\item The restriction $\phi|_{E_p} :E_p \rightarrow \{p\}\times \mathbb{R}^k$ is a linear isomorphism.
\end{enumerate}

The idea of defining vector bundle is to associate a vector space to every point in the manifold, or equivalently, parameterizing a family of vector spaces by a manifold.

\textbf{Examples:}
\begin{enumerate}
\item The tangent bundle of a smooth manifold is the disjoint union of all tangent spaces of every point in the manifold.

\item The cotangent bundle of a manifold is the disjoint union of cotangent spaces of the manifold.
\end{enumerate}

\textbf{Definition:} Let $\pi : E \rightarrow M$ be a vector bundle over M. A \textbf{section} of E is a map $F: M \rightarrow E$ such that $\pi \circ F = Id_M$. A section is called smooth if F is smooth.

\textbf{Examples:}
\begin{enumerate}
\item A section over the vector bundle of a manifold is a vector field i.e. the vector space of derivations on the manifold.

\item A section over the cotangent bundle is a covector field i.e. the vector space of one forms on the manifold.
\end{enumerate}

\subsubsection{Dolbault cohomology}
On a complex manifold, at least locally, we can define coordinates as in real manifolds, the difference is that coordinated in this case can be divided into holomorphic coordinates $z_i$ and antiholomorphic coordinates $\bar{z}_i$, for example a differential form can locally be written as 
$$\omega=f(z_1,z_2,...,z_p,\bar{z}_1,\bar{z}_2,...,\bar{z}_q)dz_1 \wedge dz_2 \wedge ... \wedge dz_p \wedge d\bar{z_1} \wedge d\bar{z}_2 \wedge ... \wedge \bar{z}_q.$$
This has an important consequence on the cohomology theory, that is the De Rham complex is now bigraded not just graded as in the real case. Thus, we have to define a holomorphic and antiholomorphic differentials, cohomologies, dimensions and so on.

\textbf{Notations:} Let M be a complex mainfold, we denote the bigraded complex vector bundle of all forms with p holomorphic and q antiholomorphic coordinate dependence i.e. of bi degree (p,q) by $\Omega^{p,q}$ and $\Gamma(\Omega^{p,q},M)$ is its space of sections.

We begin with defining adequate differential operators. Beginning with taking the exterior derivative of a generic differential form and by the chain rule, we can see that the exterior derivative can be decomposed into holomorphic and antiholomorphic parts
$$d\omega = \partial \omega + \bar{\partial}\omega$$
where $\partial$ is the holomorphic part i.e. a differential acting only on the holomorphic part of $\omega$, and $\bar{\partial}$ is the antiholomorphic part which leads to the formal definition of Dolbeault operators.

\textbf{Definition:} Let M be a complex manifold, the Dolbeault operators on its charts are defined as 
$$\partial : \Gamma(\Omega^{p,q},M) \rightarrow \Gamma(\Omega^{p+1,q},M),$$
$$\bar{\partial} : \Gamma(\Omega^{p,q},M) \rightarrow \Gamma(\Omega^{p,q+1},M).$$
such that each operator is a derivation.

The next step is to define a cohomology with respect to Dolbeault operators, defining holomorphic and antiholomorphic closed and exact forms is similar to the real case, so the general cohomology group is defined as the quotient
$$H^{p,q}(M)=\frac{ker(\bar{\partial}(\Gamma(\Omega^{p,q},M))}{Im(\bar{\partial}(\Gamma(\Omega^{p,q-1},M)))},$$
Similar to Betti numbers, the complex dimensions of Doulbeault cohomology groups are called Hodge numbers
$$h^{p,q}(M)=dim_C(H^{p,q}(M)),$$
which will be used extensively in chapter 4 and 5 in Calabi-Yau compactifications.
\subsubsection{Integrability}
\textbf{Definition:} Let M be an almost complex manifold with almost complex structure J. If the lie bracket of any two holomorphic vector fields is holomorphic, then J is said to be \textbf{integrable}. An integrable almost complex structure is said to be a \textbf{complex structure} and an even dimensional manifold M equipped with the complex structure is called a complex manifold.

\textbf{Theorem:} Let M be an almost complex manifold with almost complex structure J. J is integrable if and only if the Nijenhuis tensor defined as
$$N_J(X,Y)=[X,Y]+J[X,JY]+J[JX,Y]-[JX,JY],$$
where $[X,Y]$ is the lie bracket of the vector fields X and Y, vanishes for any two vector fields.

We now define an important type of complex manifolds namely Kahler manifolds. Firstly, we observe that to any almost complex structure J provided that the metric g is hermitian, we can associate a hermitian 2-form defined by $\omega(X,Y)=g(X,JY)$.

\textbf{Theorem:} Let M be a complex manifold with a complex structure J, a hermitian metric g and an associated hermitian 2-form $\omega$, and let $\nabla$ be the Levi-Civita connection, then the following statements are equivalent ;
\begin{enumerate}
\item $\nabla \omega=0$ 
\item $\nabla J=0$
\item $d\omega=0$
\end{enumerate}
The manifold satisfying these conditions is called a Kahler manifold.
\subsubsection{Chern classes}
The motivation to define Chern classes is the question of whether or not two vector bundles on a manifold are isomorphic. Chern classes give a necessary condition on the existence of such isomorphism i.e. if two bundles do not have equal Chern classes, they are not isomorphic.

\textbf{Definition:} Let E $\rightarrow$ M be a vector bundle over a manifold M with connection form $\nabla$ whose curvature form is $R^{\nabla}$. The total Chern class is defined as
$$c(E,t)=\det(I+\frac{i}{2\pi}R^{\nabla})$$

The total Chern class can be expanded as 
$$c(E,t)=\sum_{n=0}^M c_n(E)$$
where $M \leq max(dim(M),rk(E))$
and $c_n(E)$ is called the nth Chern class.

In this review, only the first chern class is important and is given by $c_1(E)=\frac{i}{2\pi}\text tr(R^{\nabla})$. We can see that the first chern class has a direct relation to the curvature of the manifold, and also the canonical bundle. This relation allows to define an interesting subclass of Kahler manifolds called Calabi-Yau manifolds as follows.

\textbf{Theorem:} Let M be a compact Kahler manifold of complex dimension n. The following statements are equivalent:
\begin{enumerate}
\item M has a trivial canonical bundle.
\item The holomorphic bundle $TM^{(1,0)}\rightarrow M$ has a vanishing first chern class.
\item M has holonomy SU($n/2$).
\item M is Ricci flat.
\end{enumerate}
The manifold satisfying these conditions is called Calabi-Yau manifold.\\
The theory of Calabi-Yau manifolds is presented in more details in chapter 4.

\subsection{Algebraic geometry}

In this section I will review the basic theory of sheafs and schemes. I assume the knowledge of algebraic and projective varieties, Noetherian rings, projective geometry. This section will follow Ref [5,6].

\subsubsection{Singularities and blowups}
We begin by reviewing the concepts of singularities in affine varieties and how to resolve them.

\textbf{Definition:}Let V be an irreducible variety, a point in V is called singular if the dimension of its tangent space is greater than the dimension of the variety.

In case of curves (one dimensional varieties), singular points take the form of self intersections, cusps or multiples of them.

A way to resolve a singularity i.e. replacing the singular curve with a regular one, is by blowing up the singularity. This is done by introducing additional coordinates then express the curve in terms of the new extended coordinate system which results in a regular curve and an additional exceptional curve.The procedure is as follows:

Given a curve in the coordinate system $\{x_i\}$ ; $i=1,2,...,dim(X)$ where X is the ambient space.
\begin{enumerate}
\item Introduce additional projective coordinates $\{y_i\}$ and impose the constraints $x_iy_j=x_jy_i$ for $i,j=1,2,...,dim(X)$. This defines a closed subvariety S$\subset V \times P^{n-1}$
\item Define the blowup map $\sigma :S \rightarrow V$ to be the first projection. 
\item To blowup a singularity $z\in V$ consider the map $\sigma^{-1}:V \rightarrow S$, the image of every regular point is itself while the image of z is $z \times P^{n-1}$, and the resulting curve in non singular.
\end{enumerate}
The image of the singular point is mapped to its projectified tangent space i.e. since the point has infinitely many tangents, every tangent is mapped to a point in the additional dimensions we introduced, or equivalently each tangent is mapped to a point in an exceptional curve.

\textbf{Example:} Consider the curve $y^2=x^3+x^2$, the curve has a singularity (self intersection) at $(0,0)$. To blow up the singularity, firstly we define a new projective coordinate t, then impose the condition xt=y. Substituting in the curve we get
$x^2(t^2-x-1)=0$ i.e. two curves: a non singular curve $t^2-x-1=0$ and an exceptional curve $x=0$.\\
If the resulting curve is again singular, we iterate this procedure until it is resolved. If it will take infinite number of iterations, it can be resolved by Newton's rotating ruler method which is beyond the scope of this review.

\textbf{Example:} Consider the curve $(x^2+y^2-1)^2=0$ which represents two coincident unit circles whose center is at the origin, each circle is called a branch of the curve i.e. the curve consists of two coincident branches each one of them is a unit circle whose center is at the origin. This curve has a singularity in every point since every point is an intersection point between the two branches i.e. a self intersection of the curve. This curve's singularities can not be resolved by a finite number of blowups because each singularity requires a blowup and there are infinite number of singularities. The singularities of this curve can be resolved by Newton's rotating ruler method.

\subsubsection{Sheafs}

\textbf{Definition:} Let X be a topological space. Suppose we associate a set $\mathcal{F}(\text U)$ (called a section  of the presheaf) to every open set U $\subset$ X , and a map $\rho ^{\text V}_{\text U}: \mathcal{F}(\text V) \rightarrow \mathcal{F}(\text(U)),$ for every V containing U, satisfying the following:
\begin{enumerate}
\item $\text U=\phi \implies \mathcal{F}(\text U)$ is a singleton.

\item $\forall$ open subset U of X, $\rho ^{\text U}_{\text U}$ is the identity map.

\item $\forall$ open sets $\text U \subset \text V \subset \text W$, $\rho^{\text W}_{\text U} =\rho^{\text V}_{\text U} \circ \rho^{\text W}_{\text V}$.
\end{enumerate}
This system of sets and maps is called a \textbf{presheaf}, and is denoted by $\mathcal{F}.$

If all the sets $\mathcal{F}(\text U)$ are groups, rings, fields,... ,then the presheaf $\mathcal{F}$ is called a presheaf of groups, rings, fields,... .

Note that the three conditions in the definition are exactly the condition in the definition of a functor discussed in section 1. This leads to a more general definition for a presheaf.

\textbf{Definition:} An $\mathcal{S}$-valued presheaf on a small category $\mathcal{C}$, where $\mathcal{S}$ is a category, is a contravariant functor from $\mathcal{C}$ to $\mathcal{S}$. We recover the previous definition if we set $\mathcal{C}=\mathbf{Top(X)}$(The category of open sets in the topological space X).

\textbf{Definition:} A presheaf $\mathcal{F}$ on a topological space X is a sheaf if $\forall$ open set U $\subset$ X and $\forall$ open cover of U ($\text U=\cup_{\alpha}\text U_{\alpha}$), the following conditions are satisfied:
\begin{enumerate}
\item Let $s_1$ and $s_2$ be arbitrary elements of $\mathcal{F}(\text U)$. Then, $\forall \text U_{\alpha} \rho ^{\text U}_{\text U_{\alpha}}(s_1)=  \rho^{\text U}_{\text U_{\alpha}}(s_2) \implies s_1=s_2$.
\item Let $S_{\alpha} \in \mathcal{F}(\text U)$. $\forall \text U_{\alpha},\text U_{\beta} \subset \text U$, $\rho ^{\text U_{\alpha}}_{\text U_{\alpha} {\cap} \text U_{\beta}}(s_{\alpha})=\rho ^{\text U_{\beta}}_{\text U_{\alpha}\cap \text U_{\beta}}(s_{\beta}) \implies \exists s \in \mathcal{F}(\text U)$; $s_{\alpha}=\rho ^{\text U}_{\text U_{\alpha}}$ for each $\text U_{\alpha}$.
\end{enumerate}
Basically a presheaf is a construction to associate data to open sets in a topological space, a data can be groups so the presheaf is a presheaf of groups, or rings so the presheaf is a presheaf of rings and so on. A sheaf is when this data can be "glued" together throughout the topological space, and can be restricted to smaller sets without distortion i.e. the total data equals the sum of data restricted in each subset.

\textbf{Example:} An important example of a sheaf is the presheaf of rings on Spec(A) the topological space of all prime ideals of a ring A. This sheaf is called the structure sheaf on Spec(A), and is denoted by $\mathcal{O}_{\text A}$.

\textbf{Non-Example:}  An example of a presheaf that fails to be a sheaf is the presheaf where $\mathcal{F}(\text U)$ is the set of constant maps from U to a set M. This presheaf fails to meet the second condition in the definition of a sheaf.

\subsubsection{Schemes}

\hspace{18pt}\textbf{Definition:}Let X be a topological space, and let $O_{X}$ be a sheaf of rings on X, A \textbf{Ringed space} is the pair $(X,O_X)$. In this case the sheaf $O_X$ is called the structure sheaf of X. Obviously, ringed spaces form a category.

\textbf{Definition:} A \textbf{Scheme} is a ringed space (X,$O_X$) such that every point x has a neighborhood U (called an affine neighborhood of x)satisfying the condition that the ringed space (U,$O_ X|_{\text U}$) is isomorphic to Spec(A) for some ring A.Schemes form a category 

\textbf{Examples:}
\begin{enumerate}
\item Spec(A) itself is a scheme for any ring A.
\item  Any polynomial with real coefficients determines a scheme in the real projective space, called a projective hypersurface.
\end{enumerate}

\textbf{Definition:} A scheme X with a morphism from it to another scheme S is called an \textbf{S-scheme}.

From the definition we can see that a scheme is the most general construction we can do projective geometry on. It can be considered as the algebraic counterpart of a manifold (Manifolds are locally homeomorphic to $R^n$, and a scheme is locally isomorphic to Spec(A)). We define schemes because Manifolds can not be used in algebraic geometry because here we are dealing with ratios of polynomials which is far less easier to deal with than smooth functions.

\subsubsection{Moduli}

Moduli are one of the most important objects in the theory of deformation of complex structures and consequently string theory. Physically, Moduli are massless fields with no potential, so they can take any value without requiring energy but each value can deform the complex structure to give different physical outcomes. In this review, we will deal with it in a more mathematical way.

\textbf{Definition:} A \textbf{Modulus} $\mathcal{M}$ is a contravariant functor from the category of schemes to the category of sets sending every S-scheme to the set of classes of isomorphic families of S(such a space whose points are isomorphism classes is called moduli space), and sending every morphism to its pullback.

\textbf{Example:} Consider the scheme of all annuli, two annuli are conformally equivalent (isomorphic in the scheme) is the ratio of the outer radius to the inner radius is equal in both annuli. This can be reformulated into the statement: for every positive number greater than one, there is an infinite set of equivalent annuli having the outer to inner radius ratio equal to this number. Thus, the moduli space for annuli is the set $[1,\infty]$, and the outer to inner ratio is a modulus.

\subsubsection{Fibered categories and stacks}

\subsection{Cobordisms}
Cobordisms are mathematical objects emerged to answer a question of whether a given manifold is a boundary of a higher dimensional manifold. This tool will be useful later when discussing the swampland project as cobordisms are related to global symmetries in supergravity theories as will discuss later in chapter 9 in detail. In this section we will briefly introduce cobordisms and cobordant manifolds and their category theory generalisation.

\textbf{Definition:} Let M and N be two n dimensional smooth, compact manifolds. M and N are called \textbf{cobordant} if there exists an $n+1$ dimensional manifold with boundaries W, called the cobordism joining $M$ and $N$, and two embeddings $i:M \rightarrow \partial W$, and $j:N \rightarrow \partial W$ such that 
\begin{enumerate}
    \item $i(M)\cap j(N) = \phi$
    \item $\partial W = i(M)\cup j(N)$
\end{enumerate}

\begin{figure}[h]
    \centering
    
    \includegraphics[scale=0.15]{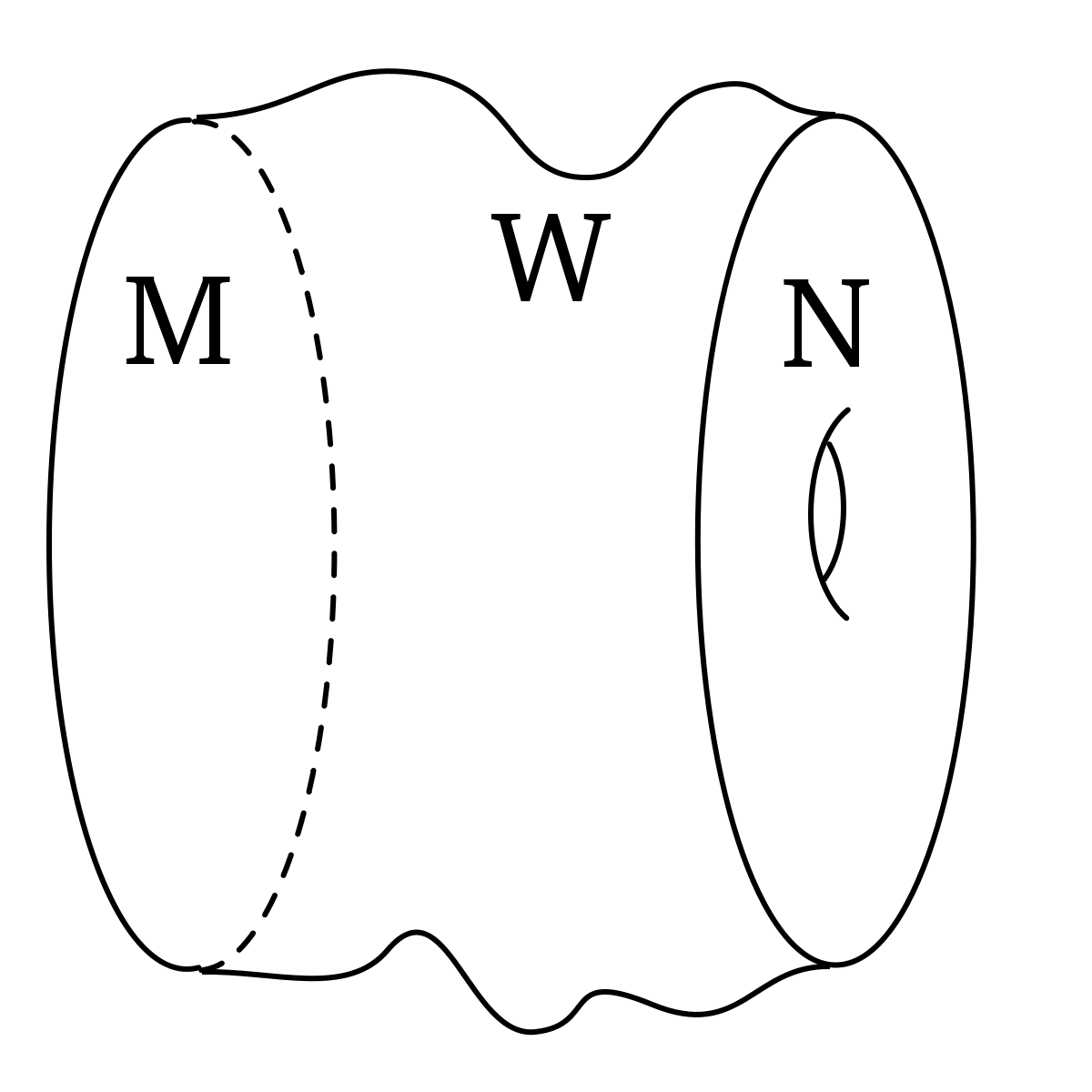}
    \caption{Here M and N are two dimensional cobordant manifolds due to the existence of the 3 dimensional manifold W whose boundaries are comprised entirely of M and N without them intersecting. }
    \label{fig:my_label}
\end{figure}

\textbf{Definition:} Let M be  n dimensional smooth compact manifold, the set of all n dimensional manifolds cobordant to M is called \textbf{the cobordism class} of M, and is denoted by [M].

Clearly the set of cobordims classes of n dimensional manifolds equipped with the disjoint union as an operation is an Abelian group (with the identity is [$\phi$] and the inverse of each element is the element itself). Thus, the set of cobordisms classes of n dimensional manifolds is also a category denoted by $\mathbf{Bord_n}$ whose objects are n dimensional manifolds and a morphism from a manifold $E$ to a manifold $F$ is the $n+1$ dimensional manifolds defining the cobordism between them.\\

\section{String compactifications}
In this chapter we review some compactification models of superstring theories without flux. As type I superstring theory is just type II superstring theory on an orbifold, it is sufficient to consider only type II and heterotic superstrings. 

\subsection{Compactification of type II string theory on a circle}
The easiest compactification scheme is to compactify on $S^1$ leading to a 9D theory. While it is not realistic, it serves as a tool to understand more sophisticated compactifications. Compactifications on circles and tori are reviewed in [7-10].

In this scheme the 10D underlying manifold of the theory is written as $\mathcal{M}_{10}= \mathcal{M}_9 \times S^1$ where $\mathcal{M}_9$ is a non compact 9D manifold.

Since one dimension is compact, its momentum must be quantized. On a circle 
$$X^9(\sigma+l,t)=X^9(\sigma , t)+2\pi R \omega$$
where $X^9$ is the coordinate along the compact dimension, $l$ is the length of the string, R is the radius of the circle, $\sigma$ and t are the parameters of the parameter space, and $\omega$ is the winding number i.e. the homotopy class of the string.

The momentum is then $p_9=\frac{k}{R}$ where k is an integer. On decomposing into left and right moving modes, the momenta are
$$p_L=\frac{k}{R}+\frac{\omega R}{\alpha'}$$
$$p_R=\frac{k}{R}-\frac{\omega R}{\alpha'}$$
with mass formulae
$$M^2_L=\frac{p_L^2}{2}+\frac{2}{\alpha'}(N_B+N_F+E_0)$$
$$M^2_R=\frac{p_R^2}{2}+\frac{2}{\alpha'}(\bar{N}_B+\bar{N}_F+\bar{E}_0)$$
where $N_B$ is the left moving bosonic mode number, $N_F$ is the left moving fermionc mode number and $E_0$ is a constant, the quantities with bars are their right moving counterparts. 
Note that there are no tachyons in the spectrum.

We now focus on the massless modes. In the original 10D theory, the space is 9D so the symmetry group is $SO(8)$, after the compactification the non compact space is 8D so the symmetry group must be $SO(7)$ i.e. we need to decompose the representations of $SO(8)$ into representations of $SO(7)$ to get the particle content after compactification.

The Neveu Schwartz (NS) sector is decomposed as $\mathbf{8}_V \rightarrow \mathbf{7} \oplus \mathbf{1}$ where $\mathbf{8}_V$ is the vector 8D representation of $SO(8)$ and $\mathbf{7}$ is the vector representation of $SO(7)$. 
For Ramond (R) sector the decompositions are $\mathbf{8}_S \rightarrow \mathbf{8}$ and $\mathbf{8}_C \rightarrow \mathbf{8}$ where $\mathbf{8}_S$ and $\mathbf{8}_C$
are the chiral representations of $SO(8)$ of left and right movers respectively, and $\mathbf{8}$ is the spinor representation of $SO(7)$. Note that all information about chirality is lost in the process because no chirality can be defined in odd dimensions, this indicates that compactification on a circle can not result in a chiral theory.

Deducing the particle content can be done by taking tensor products of NS and R sectors to form the full string contents:\\
\textbf{- NS-NS sector:}
$$(\mathbf{7}\oplus \mathbf{1}) \otimes   (\mathbf{7}\oplus \mathbf{1}) = \mathbf{27} \oplus \mathbf{21} \oplus \mathbf{7} \oplus \mathbf{7} \oplus \mathbf{1} \oplus \mathbf{1}  $$
i.e. a graviton, a KR field, two vector fields and two scalar fields.\\
\textbf{R-NS sector:}
$$\mathbf{8} \otimes (\mathbf{7} \oplus \mathbf{1})= \mathbf{48} \oplus \mathbf{8} \oplus \mathbf{8}$$
i.e. one Rarita- Schwinger field and two spinor fields. and same for NS-R sector.\\
\textbf{R-R sector:}\\
\textbf{IIA:} 
$$\mathbf{8} \otimes \mathbf{8} \rightarrow (\mathbf{1} \oplus \mathbf{7}) \oplus (\mathbf{21} \oplus \mathbf{35})$$
i.e. A scalar, one form  and a three form.\\
\textbf{IIB:}
$$\mathbf{8} \otimes \mathbf{8} \rightarrow (\mathbf{1}) \oplus (\mathbf{7} \oplus \mathbf{21}) \oplus (\mathbf{35})$$
i.e. a scalar, two form and a four form.\\
This field contents means that the theory has 32 supercharges ($\mathcal{N}=8$ supergravity in 4D) which is the maximal supersymmetry, we can then conclude that circle compactifications can not break any supersymmetry. Thus compactification on circles and spheres without flux can not give a realistic particle physics theory.

\subsection{Toroidal compactification of type II superstrings}
In this section we study the case of compactification on square tori of dimension d which are products of d circles. Firstly, we define parameters on tori. Since a torus is a product of circles and a circle can be considered as a line segment with its endpoints identified i.e. a quotient space. Therefore, we can define a torus also as a quotient space as follows:\\
Firstly, define the equivalence relation $\sim$ on $\mathbb{R}^d$ ,where d is the number of compact dimensions, by
$$X^i \sim X^i+2\pi \text R^i$$
 where $\text R^i$ is the radius of the circle in the i-th compact dimension comprising the torus and i runs over the compact dimensions.
 
 A torus is defined as the quotient space $\mathbb{R}^d/\sim$.
 In the case that all $\text R^i$ are equal the torus is called a square torus.\\
 By this definition we have d compact dimensions, 
 this means that there are d quantized momenta $\text P_i=\frac{\text k+i}{\text R}$ where $k_i$ are integers. Additionally, in this case we have winding number $\omega_i$ for every compactified dimension measuring how many times the string is wrapped on it, this number coincides with the homology class of a circle. 
 
 The Lagrangian for this theory is given by
 $$S=\frac{1}{2\pi} \int_0^{\infty} d^2\sigma  [\frac{1}{2\alpha'} G_{ij}(\gamma^{\mu \nu}\partial_{\mu}X^i\partial_{\nu}X^j)   +    \frac{1}{\alpha'}B_{ij}\partial_{\sigma_1}X^i\partial_{\sigma_2}X^j], $$
where $G_{ij}$ is a symmetric 2 form (graviton), $B_{ij}$ is an antisymmetric 2 form (KR field), $\sigma_1$, $\sigma_2$ are the parameters of the parameter space of the worldsheet, $X^i$ are the coordinates on the worldsheet and $\gamma^{\mu \nu}=diag[1,-1]$.

Since this compactification does not alter any oscillatory mode (because it is already periodic), it is sufficient to work on the zero modes, i.e. we can expand the coordinates as follows
$$X^i=x_0^i+\dot{x}^i\sigma_1+\frac{2\pi \text R}{\text l}\omega^i \sigma_2,$$
where l is the length of the string. Here we set $\sigma_1$ to be the time coordinate and $\sigma_2$ to be the parameter on the string at a fixed time.\\
Substituting in the Lagrangian and integrating, we get
$$S=\frac{\text l}{2\pi}[\frac{1}{2\alpha'}G_{ij}(\dot{x}^i \dot{x}^j - (\frac{2 \pi \text R}{\text l})^2 \omega^i \omega^j)+\frac{2\pi \text R}{\text l \alpha'}B_{ij}\dot{x}^i \omega^j]$$
The canonical momentum is then given by
$$p_i=\frac{\text l}{2 \pi \text R}G_{ij}\dot{x}^j+B_{ij}\omega^j,$$
But the quantization condition is $p_i=\frac{k_i}{R}$, equating and solving for $\dot{x}^j$ we get
$$\dot{x}^j=\frac{2 \pi \text R}{\text l}G^{ij}(\frac{k_i}{\text R}-K_{ik}\omega^k)$$
substituting back to get the worldsheet coordinates and after some algebra we get the left and right moving sectors of the momentum
$$p_{L,i}=\frac{k_i}{\text R}+\frac{\text R\omega^j}{\alpha'}(G_{ij}-B_{ij}),$$
$$p_{\text R,i}=\frac{k_i}{\text R}+\frac{\text R\omega^j}{\alpha'}(-G_{ij}-B_{ij}).$$
Note that the momenta are no longer parameterized by one integers but two. Thus, they form a 2D lattice where D is the dimension of the compact torus used. This lattice is even, self dual and is called Narain lattice.

To prove this, first define the coordinates of the lattice to be $(p_L,p_R)$.
In the Lorentzian signature scalar product, the product of any two points in the lattice is even.\\
Proof: Let $(p_L,p_R)$ and $(p'_L,p'_R)$ be two arbitrary points in that lattice, the inner product is
$$(p_L,p_R)\cdot(p_L,p_R)=2\sum_i[k^i\omega'_i+k'_i \omega^i]$$
which is an even integer.

The self dual part is obvious because the theory must be modular invariant which is a well known symmetry that holds in string theory.

The moduli space of the theory can be deduced from this lattice description (equivalent lattices correspond to equivalent theories). Narain lattices by definition are unique up to $SO(d,d)$ transformations, but in this case each point in the lattice used here is invariant under $SO(d) \times SO(d)$ because mass states of the theory depends only on the sum of squares of momenta as will deduced afterwards. Additionally, two lattices are give the same theory if they are T dual i.e. invariant under T duality group $SO(d,d,Z)$ (Remember that T duality change the sign of the right movers only so the point in the lattice will be transformed by an integer value). From this we can say that the moduli space of the theory is the quotient group
$$\frac{SO(d,d)}{SO(d)\times SO(d) \times SO(d,d,Z)}.$$

Now we deduce the field content of the theory: From the formulae for momenta deduced earlier, we can derive the mass formulae for left and right movers to be
$$M_L^2=\frac{2}{\alpha'}(N_F+N_B+E_0)+\frac{p^2_L}{2},$$
$$M_R^2=\frac{2}{\alpha'}(\bar{N}_F+\bar{N}_B+E_0)+\frac{p^2_R}{2}.$$
The field content is derived from the level matching condition as in the usual string theory. Here we only stress on the fact that the field content has 32 supercharges i.e. $\mathcal{N}=8$ supergravity in 4D, this is because tori are just products of circles and we saw in the previous section that compactifications on circles can not break any supersymmetry, thus, the theory will be maximally supersymmetric.

\subsection{Compactification of type IIB superstrings on $AdS_5 \times S^5$}
Compactification on $AdS_5 \times S^5$ particularly is an interesting model because it was proven to be exactly dual to $\mathcal{N}=4$ super Yang Mills theory by two independent methods: the first method was by using AdS/CFT correspondence, and was shown to be the case by Metsaev and Tseytlin [11], and the second was by considering a stack of N coincident D3 branes in the limit $\text N \rightarrow \infty$ [12] and studied further in [13]. This means that the field content and the supersymmetry of the theory is already known. Here we review the action, equations of motion and some properties.

Firstly, note that the symmetry group is the super lie group $PSU(2,2|4)$ whose super lie algebra has no representations in terms of supermatrices but in terms of quotients of supermatrices, for more mathematical properties see [14-16].

The action is unique and consists of two terms: the metric term and the Wess Zumino term. The metric term is as usual
$$S_1=-\frac{\text R}{4 \pi \alpha'} \int d^2 \sigma [\sqrt{-h}h^{\mu \nu}G_{\mu \nu}],$$
where R is the radius of the metric of $AdS_5 \times S^5$, $h^{\mu \nu}$ is the inverse of the worldsheet metric. It can also be written in terms of a one form $J_1$ defined in [17] as
$$S_1=\frac{\sqrt{\lambda}}{16 \pi} \int str(J_1 \wedge \star J_1),$$ 
where $\lambda = \frac{\text R^4}{\alpha'^2}$, str is the supertrace, and $\star J_1$ is the Hodge dual to $J_1$.

The Wess Zumino term is proportional to the only non zero $psu(2,2|4)$ invariant two form namely $str(J_2 \wedge J_3)$. Thus, the total action is given by
$$S=\frac{\sqrt{\lambda}}{16 \pi} \int str(J_1 \wedge \star J_1)-\frac{\sqrt{\lambda}}{8 \pi} \int str(J_2 \wedge J_3).$$
The invariance under $PSU(2,2|4)$ induces a conserved current which is given by 
$$J=J_1+\star J_3.$$
The equations of motion coincide with the conservation of the dual of the current namely $d\star J=0$, in terms of $J_i$ we get
$$d\star J_1+ dJ_3=0.$$
The general features of this theory is that its bosonic part coincides with the standard sigma model on $AdS_5 \times S^5$, the theory has local $\kappa$ symmetry which is responsible of removing half of the supersymmetries, that is why this model has $\mathcal{N}=4$ supersymmetry instead of $\mathcal{N}=8$. The analysis of kappa symmetry is beyond the scope of this review for more details see [18]. Another feature is that the theory reduces to the usual type IIB superstring theory in the limit $\text R \rightarrow \infty$. For more detailed analysis of these type of models see [19-21].

Other compactifications to $AdS$ space exist and are numerous, for example there are compactifications on orbifolds to $AdS$ like $AdS_5 \times S^5/Z_2$ but this type of models were rules out due to Dirac quantization rule of charges which will be discussed in chapter 7.
Other models include compactifications on various combinations of circles, tori and other manifolds are reviewed in [22] and references therein.

\subsection{Compactification of heterotic strings on a circle}
In this section we review the compactification of heterotic superstrings on a circle without Wilson lines, compactifications with Wilson lines are presented in [9].\\
Heterotic strings are combination of bosonic left movers and superstrings as right movers. Thus, we repeat the analysis in section 3.2 with the adequate changes and show that using heterotic string theory can eliminate half of supercharges in the same compactification model.

Once again, compactification on a circle means that one coordinate must satisfy
$$X^{9}(\sigma+\text l,t)=X^{9}(\sigma,t)+2\pi R \omega,$$
where $X^9$ is the compact dimension, $\sigma$ and $\tau$ are the parameters parameterizing the string's worldsheet, l is the length of the string, R is the radius of the circle and $\omega$ is the winding number.

The corresponding momentum is quantized as in type II case with the difference that the left moving part is bosonic and is 26 dimensional. Thus, the left moving momentum consists of two parts: 10D part similar to the superstring momentum discussed before, and an additional internal 16D momentum denoted by $P_{int}=\sqrt{\frac{2}{\alpha'}}P$ where the $\sqrt{\frac{2}{\alpha'}}$ factor is added for future convenience.

The momenta of the 10D part are given by
$$p_R=\frac{k}{R}+\frac{R \omega}{\alpha'},$$
$$p_L=\frac{k}{R}-\frac{R \omega}{\alpha'},$$
where $p_R$ and $p_L $is the right and the left moving momenta respectively and
the square of the mass is given by
$$M^2_R=\frac{p^2_R}{2}+\frac{2}{\alpha'}(\bar{N_B}+\bar{N_F}+E_0),$$
$$M^2_L=\frac{p^2_L}{2}+\frac{P^2_{int}}{2}+\frac{2}{\alpha'}(N_B-1)=\frac{p^2_L}{2}+\frac{2}{\alpha'}(N_B+P-1),$$
where $M_R$ and $M_L$ are the right and left movers' mass respectively.\\
Note that in the case of bosonic strings the momentum shift is known to be -1 which is the reason why bosonic string spectrum has tachyons, in cases of heterotic $E_8 \times E_8$ and SO(32) strings tachyons do not exist. Also, the left mover's mass has a contribution from the internal momentum but no fermionic contribution as it is a bosonic sector.

We now derive the particle content and prove that this theory has half the number of supercharges of its type II counterpart.
The decomposition of $SO(8)$ for right movers is done as before namely, $\mathbf{8}_V \rightarrow \mathbf{7} \oplus \mathbf{1}$ and $\mathbf{8}_C \rightarrow \mathbf{8}$. As before, chirality is lost i.e. it is not possible to obtain chiral theories from this compactification scheme. For the left movers we do not have Ramond and Neveu-Schwartz sectors, the lowest energy particle states are as follows:
\begin{enumerate}
\item $\alpha_{-1}^i|0>$ where $i-1,2,...,10$ are the 10D states propagating with the right moving superstring sector and is decomposed as $\alpha_{-1}^i|0> \rightarrow \mathbf{7} \oplus \mathbf{1}$.
\item $\alpha^I_{-1}|0>$ where $I=11,12,...,26$ are the 16D states in the internal space of the bosonic sector which decompose as $\alpha^I_{-1}|0> \rightarrow 1$, as they are in a different space so no gauge group affects them.
\item $|P>_I$ which are the on shell momentum states (remember that bosonic string states are specified by two parameters: energy and momentum), these states also decomposed as $|P>_I \rightarrow \mathbf{1}$ as it is in the internal space.
\end{enumerate}

To get the whole particle content of the heterotic theory, we take tensor products of each superstring state with each bosonic string state
$$\mathbf{8}_V \otimes \alpha^i_{-1}|0> \rightarrow (\mathbf{7}\oplus \mathbf{1})\otimes (\mathbf{7}\oplus \mathbf{1}) = \mathbf{27} \oplus \mathbf{21} \oplus \mathbf{7} \oplus \mathbf{7} \oplus \mathbf{1} \oplus \mathbf{1}$$
i.e. a metric, a KR field, two vectors , a scalar and a dilaton.\\
$$\mathbf{8}_V \otimes \alpha_{-1}^I|0> \rightarrow (\mathbf{7}\oplus \mathbf{1}) \otimes \mathbf{1}= \mathbf{7} \oplus \mathbf{1}$$
i.e. a spinor and a scalar.\\ 
$$\mathbf{8}_V \otimes |P>_I \rightarrow (\mathbf{7} \oplus \mathbf{1}) \times \mathbf{1} = \mathbf{7} \oplus \mathbf{1}$$
i.e. a vector and a scalar.\\
$$\mathbf{8}_C \otimes \alpha_{-1}^i|0> \rightarrow \mathbf{8} \otimes( \mathbf{7} \oplus \mathbf{1})= \mathbf{48} \oplus \mathbf{8} \oplus \mathbf{8}$$
i.e. a Rarita-Schwinger field (gravitino) and two spinors.\\
$$\mathbf{8}_C \otimes \alpha^I_{-1}|0> \rightarrow \mathbf{8} \otimes \mathbf{1}=\mathbf{8}$$
i.e. a spinor.\\
$$\mathbf{8}_C \otimes |P>_I \rightarrow \mathbf{8} \otimes \mathbf{1}=\mathbf{8}$$
i.e. a spinor.

This particle content is a 9D supergravity with 16 supercharges so in 4D we get $\mathcal{N}=4$ supergravity, this is half the supersymmetry of type II on a circle due to the fact that bosonic states never decomposes to spinors, this results in decreasing the number of spinors, so the number of supercharges by half.

\subsection{Toroidal compactification of heterotic superstrings}
We generalize the previous section to tori (products of circles). As in the case of type II superstrings we consider square d dimensional tori.\\
The action of the theory is a mixture between actions on the 10D space of superstrings and the internal bosonic 16D space and is given by
$$S=\int d^2\sigma [G_{IJ} \eta^{\mu\nu}-B_{IJ}\epsilon^{\mu\nu}]\partial_{\mu}X^I\partial_{\nu}X^J,$$
where $I,J$ are indices in the internal space, $\mu,\nu$ are indices on the superstring space, $\epsilon^{\mu\nu}$ is the totally antisymmetric tensor in the superstring space, $\eta^{\mu\nu}$ is the inverse of the Minkowski metric in the superstring space, $\sigma_1$ and $\sigma_2$ are the parameters in the parameter space and the derivatives are on superstring space coordinates.

Following similar calculations to section 3.3, the momenta are given by
$$P_I=\sqrt{\frac{2}{\alpha'}}P_I$$
$$p_{L,i}=\frac{k_i}{R}+\frac{R\omega^j}{\alpha'}(G_{ij}-B_{ij}),$$
$$p_{R,i}=\frac{k_i}{R}+\frac{R\omega^j}{\alpha'}(-G_{ij}-B_{ij}),$$
where $P_I$ is the internal momentum in the 16D internal space of the left movers and the index I runs over this space and $\omega^i$ are the winding numbers.\\
These results are similar to that of type II but with extra internal momentum for the bosonic extra 16D internal space. The extra momentum is a part of the left moving sector i.e. the quantized left moving momentum space is $16+d$ dimensional not d dimensional as the right moving momentum space.

The next step is to deduce the moduli space of the theory. The momenta form a $d(d+16)$ Narain lattice with coordinates represents left and right moving momenta (so the dimension of the lattice is the product of dimensionalities of the two momentum spaces). 

As the mass spectrum only depends on the magnitude of vectors in the lattice, the lattice is unique up to $SO(16+d,d)$ rotations, but left and right moving momenta rotate in the groups $SO(16+d)$ and $SO(d)$ respectively i.e. each point is invariant under the group $SO(16+d) \times SO(d)$. Considering also the T duality group $SO(16-d,d,Z)$, we get the moduli space 
$$\frac{SO(16+d,d)}{SO(16_d)\times SO(d) \times SO(16-d,d,Z)}.$$
Finally, the mass spectrum is given by
$$M_L^2=\frac{2}{\alpha'}(P_IP^I+N_B-1)+\frac{p^2_L}{2},$$
$$M_R^2=\frac{2}{\alpha'}(\bar{N}_F+\bar{N}_B+E_0)+\frac{p^2_R}{2},$$
here the index I is being summed over.

The particle content of the theory is deduced similar to the last section, it gives a 9D supergravity with 16 supercharges i.e. $\mathcal{N}=4$, 4D supergravity which is half of the number of supercharges of type II toroidal compactification. 

It is worth mentioning that the resulting theory is not chiral (no chiral fermions can be found in the spectrum) because on decomposing group representations, chiral representations with different chiralities decompose to the same lower dimensional group, so the chirality is lost, this happens in circle and toroidal compactifications because only rotation groups are used.

\subsection{Motivation for Calabi-Yau compactifications}
In all the previous compactifications, there are some problems:
\begin{enumerate}
\item There are too many supercharges; a realistic model of particle physics must be at most $\mathcal{N}=1$ supersymmetric in 4D which is not true in circle, toroidal compactifications and compactifications to $AdS$ spaces discussed before.
\item The models discussed before can never produce chiral fermions which were observed in nature (this can also be done in some orbifold compactifications [23-25] but it is beyond the scope of the review).
\end{enumerate}
To solve these problems we must choose another suitable manifold to compactify on. In their renowned paper [26] Candelas, Horowitz, Strominger and Witten showed that this manifold must be a Calabi-Yau manifold.

The paper studies heterotic string compactification on Calabi-Yau manifolds. They proved that $\mathcal{N}=1$ 4D supersymmetry requires the manifold to have $SU(3)$ holonomy, and it must be Ricci flat to eliminate the conformal anomaly as we will discuss in chapter 5.

Other results show that Calabi-Yau manifold compactifications can produce chiral fermions and generations of particles similar to the standard model (the number of generations is equal to half of the manifold's Euler characteristic). Thus, it is the best candidate to compactify on without turning on any flux.

Note that in any of the previous compactification schemes turning on flux can break supersymmetry and introduce chiral fermions which gives us a realistic models without dealing with complicated manifolds like Calabi-Yau manifolds, flux compactifications are discussed in detail in chapter 7.

\section{Calabi-Yau compactifications}
In this chapter we show why compactifications on Calabi-Yau manifolds without flux give the most phenomenologically plausible low energy particle physics theories than compactifications any other manifolds (also without flux). We also discuss the problems this scheme faced resulting in the introduction of flux compactifications and the use of new mathematics namely, generalised complex geometry.

We study the aforementioned topics in the cases of type II superstrings (type I theories can be viewed as type II theories on an orbifold), heterotic superstrings and heterotic non supersymmetric string theories. Then, we briefly review some explicit models on specific Calabi-Yau manifolds.\\

\subsection{Compactifications of bosonic strings as a toy model}
In this section we present the technique used in [27] and use it on bosonic string theory as a toy model so that we can quote its results in the later sections.\\
We begin with the Polyakov action in D dimensions describing a string propagating on an ambient manifold $\mathcal{M}$:
$$S=-\frac{1}{4\pi \alpha'}\int d^2\sigma \sqrt{\gamma}[ (\gamma^{\alpha \beta} G_{MN}(X)+i\epsilon^{\alpha \beta} B_{MN}(X))\partial_{\alpha} X^M \partial_{\beta} X^N+\alpha'\Phi \text R],$$
where $M,N=1,2,...,D$, $\alpha'$ is the Regge slope, $\alpha,\beta=1,2$, $d^2\sigma=d\sigma^1 d\sigma^2$ such that $\sigma^1$ and $\sigma^2$ are the parameters used to parameterize the world sheet, $\gamma^{\alpha \beta}$ is the induced metric on the world sheet, $\gamma=\det{\gamma^{\alpha \beta}}$, $G_{MN}$ is the metric of the ambient manifold, $\epsilon^{\alpha \beta}$ is the totally antisymmetric tensor i.e. the permutation symbol, $B_{MN}$ is an antisymmetric tensor field i.e. Kalb-Ramond field, $\partial_{\alpha}=\frac{\partial}{\partial \sigma^{\alpha}}$, $X^M$ are the coordinates on the ambient manifold, $\Phi$ is a scalar field i.e. the dilaton and R is the Ricci scalar of the world sheet.

In this chapter we use complex coordinates as parameters i.e. $\sigma^1=z$ and $\sigma^2=\bar{z}$ because they are more convenient in the later transformation to superstring models. In this case we use the notations $\partial_1=\partial=\partial_z$ and $\partial_2=\bar{\partial}=\partial_{\bar{z}}$. We also use the conformal gauge $\gamma^{\alpha \beta}= e^{-2\eta}\delta^{\alpha \beta}$, where $\eta$ is a function of the parameters $\sigma^1$ and $\sigma^2$. In this case $R=-4 \partial \bar{\partial} \eta e^{-2\eta}$.\\
The action in this gauge reads
\begin{equation}
S=-\frac{1}{4\pi \alpha'}\int d^2z e^{(d-2)\eta}[ (G_{MN}(X)+i B_{MN}(X))\partial_{\alpha} X^M \partial_{\beta} X^N+\alpha'\Phi (-4\partial \bar{\partial}\eta)],
\end{equation}
where $d$ is the dimension of the target space, but since in string theory the target space is the world sheet i.e. 2 dimensional therefore, the exponential factor cancels from the action.

For this theory to be conformally invariant i.e. conformal anomaly free, the expectation value of the trace of the energy momentum tensor must vanish. Calculating the mentioned quantity we get
$$<T^{\rho}_{\rho}>=(-\frac{1}{2\alpha'}\beta^{G}_{MN}G^{\alpha\beta}-\frac{i}{2\alpha'}\beta^B_{MN}\epsilon^{\alpha \beta}) \partial_{\alpha}X^M\partial_{\beta}X^N+2\beta^{\Phi}\partial \bar{\partial}\eta e^{-2\eta},$$
where $\rho=1,2$, $G^{\alpha \beta}$ is the pulled back metric to the world sheet, and the beta functions $\beta^G$, $\beta^B$ and $\beta^{\Phi}$ are given by
\begin{equation}
    \beta^G_{MN}=\alpha' (R_{MN}+2\nabla_M \nabla_N \Phi-\frac{1}{4}H_{MPQ}H_N^{PQ})+O(\alpha'^2),
\end{equation}
\begin{equation}
    \beta^B_{MN}=\alpha'(-\frac{1}{2}\nabla^PH_{PMN}+\nabla^P\Phi H_{PMN})+O(\alpha'^2),
\end{equation}
\begin{equation}
    \beta^{\Phi}=\alpha'(\frac{D-26}{6\alpha'}-\frac{1}{2}\nabla^2\Phi + \nabla_M\Phi \nabla^M \Phi - \frac{1}{24}H_{MNP}H^{MNP})+O(\alpha'^2),
\end{equation}
where $H=dB$ is the field strength of the field $B$, $\nabla_M$ is the covariant derivative, $D$ is the dimensions of the ambient manifold and $R_{MN}$ is the Ricci tensor of the ambient manifold. For the details of the calculations see [28-31].

We require that $<T^{\rho}_{\rho}>=0$ i.e. all the beta functions must be zero for the theory to be anomaly free. Turning off all flux i.e. $H=0$ we get the requirements
\begin{equation}
R_{MN}+2\nabla_M\nabla_N\Phi=0,
\end{equation}
\begin{equation}
\frac{D-26}{6\alpha'}-\frac{1}{2}\nabla^2\Phi + \nabla_M \Phi \nabla^M \Phi =0.
\end{equation}
So far we were using a general D dimensional manifold as an ambient manifold for the theory and we need to compactify all the extra dimensions to get a 4 dimensional particle physics theory i.e. the ambient manifold must be decomposed as $\mathcal{M}_D=\mathcal{M}_4 \times \mathcal{M}_{D-4}$, where $\mathcal{M}_4$ is a 4 dimensional manifold called the external manifold which we take to be Minkowskian, and $\mathcal{M}_{D-4}$ is a (D-4) dimensional manifold called the internal manifold, and we require that the internal manifold to be compact. Also, we impose the condition $\nabla^2\Phi=0$ otherwise we will have a fifth long range force which is not observed in nature. This implies that $\nabla_M\Phi$ is a constant and since the internal manifold is compact $\Phi$ must attain a maximum on $\mathcal{M}_{D-4}$ i.e. $\nabla_M\Phi=0$ at this point so on the whole manifold. Using these requirements in equations 4 and 5, we get
$$R_{ab}=0,$$
$$D=26$$
where $a,b$ run over the internal manifold's coordinated, this means that the internal manifold is Ricci flat and the spacetime dimensions must be 26.

Since bosonic string theory has no fermions in its spectrum therefore, it has no spinor fields so the internal manifold may not be Calabi-Yau. However, bosonic string theory can not be used to construct a realistic model of particle physics at low enegies but used just as a toy model for calculations needed later as we will see in the next few sections.
\subsection{Type II superstring theories compactifications on Calabi-Yau manifolds}
In this section we will repeat the analysis in the previous section in the case of type II syperstring theories.\\
The first step is to introduce superpartners for the bosonic variables. We define the Rarita-Schwiger field $\psi^M$ to be the superpartner of $G_{MN}$ called a gravitino, and $\chi^M$ to be the superpartner of $X^M$ called the gaugino.

The next step is to define an action which is invariant under local supersymmetry transformations. To do this step we need to define new variables, first of all the introduction of a fermionic variables implies the introduction of fermionic coordinates on the world sheet, we denote them by $\theta$ and $\bar{\theta}$. Fermionic coordinates are Grassmannian numbers i.e. they satisfy $\theta^2=\bar{\theta}^2=0$, the theory of Grassmannian numbers is beyond the scope of this review, for more details see [32]. Accordingly, we define a new derivative operators as follows
$$D=\partial_{\theta}+\theta \partial_z.$$
$$\bar{D}=\partial_{\bar{\theta}}+\bar{\theta}\partial_{\bar{z}}.$$
where $\partial_{\theta}=\frac{\partial}{\partial \theta}$ and $\partial_{\bar{\theta}}=\frac{\partial}{\partial\bar{\theta}}$.\\
We decompose the superfields according to Grassmannian numbers theory as follows:
$$X^M(z,\bar{z},\theta,\bar{\theta})=X^M(z,\bar{z})+\theta \chi^M(z) + \bar{\theta} \bar{\chi^M}(\bar{z}),$$ 
$$G_{MN}(z,\bar{z},\theta,\bar{\theta})=G_{MN}+\theta \psi^M(z)+\bar{\theta}\bar{\psi}(\bar{z}).$$
Noting that in the complex coordinates and their Grassmannian counterparts all gamma matrices vanish except $\gamma^z_{\theta \theta}$ and $\gamma^{\bar{z}}_{\bar{\theta}\bar{\theta}}$, and both are equal to the identity matrix. We can use the action (1) directly with the following transformations
$$X^M \rightarrow X^M(z,\bar{z},\theta ,\bar{\theta}),$$
$$G_{MN} \rightarrow G_{MN}(z,\bar{z}, \theta ,\bar{\theta}),$$
$$\partial \rightarrow D, \ \ \ \ \ \ \ \ \ \bar{\partial} \rightarrow \bar{D},$$
$$\eta \rightarrow \eta(z,\bar{z}.\theta ,\bar{\theta}),$$
$$d^2z \rightarrow d^2zd^2\theta,$$
where $d^2\theta = d\theta d\bar{\theta}$.\\
The action then reads

   $$ S=-\frac{1}{4\pi \alpha'}\int d^2z d^2\theta [(G_{MN}(z,\bar{z},\theta,\bar{\theta})+i B_{MN}(z,\bar{z},\theta,\bar{\theta})))D X^M(z,\bar{z},\theta,\bar{\theta})) \bar{D} X^N(z,\bar{z},\theta,\bar{\theta}))$$
    \begin{equation}
    +\alpha' \Phi (-4D \bar{D}\eta(z,\bar{z},\theta,\bar{\theta})))].
\end{equation}
This action is clearly invariant under supersymmetry transformations. For the explicit form of the transformations (without flux) see [30].

With similar calculations to the previous section, the expectation value of the energy momentum tensor is given by
$$<T^{\rho}_{\rho}>=(-\frac{1}{2\alpha'}\beta^{G}_{MN}G^{\alpha\beta}-\frac{i}{2\alpha'}\beta^B_{MN}\epsilon^{\alpha \beta}) D X^M \bar{D}X^N +2\beta^{\Phi}D \bar{D}\eta e^{-2\eta},$$
where the beta function are given by
\begin{equation}
    \beta^G_{MN}=\alpha' (R_{MN}+2\nabla_M \nabla_N \Phi-\frac{1}{4}H_{MPQ}H_N^{PQ})+O(\alpha'^2),
\end{equation}
\begin{equation}
    \beta^B_{MN}=\alpha'(-\frac{1}{2}\nabla^PH_{PMN}+\nabla^P\Phi H_{PMN})+O(\alpha'^2),
\end{equation}
\begin{equation}
    \beta^{\Phi}=\alpha'(\frac{D-10}{4\alpha'}-\frac{1}{2}\nabla^2\Phi + \nabla_M\Phi \nabla^M \Phi - \frac{1}{24}H_{MNP}H^{MNP})+O(\alpha'^2),
\end{equation}
where $H=dB$ is the field strength of the field $B$, $\nabla_M$ is the covariant derivative but replacing the ordinary derivative with the previously introduced supersymmetric derivative, $D$ is the dimensions of the ambient manifold and $R_{MN}$ is the Ricci tensor of the ambient manifold.

Similar to the previous section we turn off all fluxes and require that all the beta functions to vanish. This gives us the conditions
$$D=10,$$
$$R_{MN}=0,$$
i.e. the theory is anomaly free only in 10 spacetime dimensions, and the ambient manifold is Ricci flat.

The next step is to assume that the ambient manifold is of the form $\mathcal{M}_4\times \mathcal{M}_6$, where $\mathcal{M}_4$ is Minkowski space. The condition on the internal space is
$$R_{mn}=0,$$
where $m,n=1,2,...,6$.

On the other hand the supersymmetry transformation for $\psi^M$ can be used to derive another condition on the internal manifold as follows:\\
The transformation is
$$\delta \psi^M= \nabla_M \epsilon,$$
where $\epsilon$ is the supersymmetry parameter and a similar condition for $\bar{\psi}$, we continue the analysis with $\psi$ only and it should be understood that there are similar equations for $\bar{\psi}$.

Since the action is supersymmetric therefore, $\delta \psi^M=\delta \bar{\psi}=0$. Thus
\begin{equation}
  \nabla_M\epsilon=0.  
\end{equation}
And since the ambient manifold was decomposed into internal and external manifolds, the spinor $\epsilon$ must also be decomposed to an internal part and an external part
$$\epsilon=\xi \otimes \rho.$$
where $\xi$ is a spinor in the external Minkowsli space and $\rho$ is a spinor in the internal manifold.\\
Substituting in eq.(11) and noting that the covariant derivative vanish on the Minkowski space, we get the condition
$$\nabla_m \rho =0.$$
This means that there exist two covariantly constant spinors on the internal manifold(one from $\psi$ and one from $\bar{\psi}$).
Thus, The internal manifold is a 6D complex manifold with two covariantly constant spinors which is Ricci scalar i.e. a Calabi Yau manifold.\\
The problem with this result is that there are two covariantly constant spinors on the manifold, this means that the resulted particle physics theory in 4D will be $\mathcal{N}=2$ supersymmetric which is not plausible phenomenologically as discussed before in chapter 4.

Another concern is that Calabi Yau manifolds have large moduli spaces which are translated into the spectrum as massless fields with no potential. This means that the vacuum expectation value of these fields can change  without any energy cost i.e. freely, this kind of fields was not observed in nature moreover, the masses of particles and values of coupling constants in the 4D theory depend on the vacuum expectation value of the moduli. Thus, we need to fix moduli values, this process is called moduli stabilization and require some flux to be turned on motivating new type of compactifications called flux compactifications which will be discussed in chapter 7.

\subsection{Heterotic superstring theories compactifications on Calabi-Yau manifolds}
In this section we repeat the procedure in the previous section in the case of heterotic superstring theories.

The superpartners are to be introduced similar to the previous section but in this case we introduce one Grassmannian coordinate $\theta$, that is because in heterotic string theory only the left movers are supersymmetric, we also use complex coordinates.
To define an appropriate action we begin with defining new supersymmetric values similar to the previous section. We define a derivative operator
$$D=\partial_{\theta}+\theta \partial_z,$$
and supersymmetric variables as follows:
$$X^M(z,\bar{z},\theta)=X^M(z,\bar{z})+\theta \chi^M(z),$$ 
$$G_{MN}(z,\bar{z},\theta,\bar{\theta})=G_{MN}+\theta \psi^M(z).$$
The action is defined using (1) with the following transformations 
$$X^M \rightarrow X^M(z,\bar{z},\theta ,\bar(\theta)),$$
$$G_{MN} \rightarrow G_{MN}(z,\bar{z}, \theta),$$
$$\partial \rightarrow D,$$
$$\eta \rightarrow \eta(z,\bar{z},\theta),$$
$$d^2z \rightarrow d^2zd\theta,$$
The action then reads
   $$ S=-\frac{1}{4\pi \alpha'}\int d^2z d\theta [(G_{MN}(z,\bar{z},\theta)+i B_{MN}(z,\bar{z},\theta)))D X^M(z,\bar{z},\theta)) \bar{\partial} X^N(z,\bar{z},\theta))$$
    \begin{equation}
    +\alpha' \Phi (-4D \bar{\partial}\eta(z,\bar{z},\theta)))].
\end{equation}
The expectation value of the energy momentum tensor is given by
$$<T^{\rho}_{\rho}>=(-\frac{1}{2\alpha'}\beta^{G}_{MN}G^{\alpha\beta}-\frac{i}{2\alpha'}\beta^B_{MN}\epsilon^{\alpha \beta}) D X^M \bar{\partial}X^N +2\beta^{\Phi}D \bar{\partial}\eta e^{-2\eta},$$
where the beta function are similar to the beta functions in equations 8,9,10 but with only half of the derivatives are supersymmetric. Equating them to zero we get
$$D=10,$$
$$R_{MN}=0.$$
i.e. the theory only anomaly free in 10 spacetime dimensions and the ambient manifold is again Ricci flat.\\
Decomposing the manifold into an external and internal part $\mathcal{M}=\mathcal{M}_4 \times \mathcal{M}_{6}$, where $\mathcal{M}_4$ is a Minkowski space and $\mathcal{M}_6$ is compact, the condition on the internal manifold is
$$R_{mn}=0.$$
The transformation is
$$\delta \psi^M= \nabla_M \epsilon,$$
where $\epsilon$ is the supersymmetry parameter.

The action is supersymmetric therefore, $\delta \psi^M=0$. Thus
\begin{equation}
  \nabla_M\epsilon=0.  
\end{equation}
The ambient manifold was decomposed into internal and external manifolds thus, the spinor $\epsilon$ must also be decomposed to an internal part and an external part
$$\epsilon=\xi \otimes \rho.$$
where $\xi$ is a spinor in the external Minkowsli space and $\rho$ is a spinor in the internal manifold.\\
Substituting in eq.(13) we get the condition
$$\nabla_m \rho =0.$$

 On the contrary of the previous section, here there exists only one covariantly constant spinors on the internal manifold.
Thus, The internal manifold is a Calabi-Yau manifold.

This compactification scheme gives $\mathcal{N}=1$ supersymmetric 4D theory which is phenomenologically plausible but the issue of moduli is still a problem for this scheme as well.
\subsection{Non-supersymmetric string theory compactifications on Calabi-Yau manifolds}
Non supersymmetric string theories are theories resulting from supersymmetry breaking of heterotic superstring theories, the most famous non supersymmetric string theory  is the $SO(16)\times SO(16)$ which is the theory resulted from supersymmetry breaking of $E_8\times E_8$ heterotic string theory, or equivalently its T-dual the $SO(32)$ heterotic string theory. The details of non supersymmetric string theories are beyond the scope of this review, for more details see [33].

We will not repeat the procedure again, instead we describe how to get non supersymmetric string theories compactifications from heterotic string theories compactifications. The relation between the two type of theories' compactifications is shown in fig.5 below.
To complete our description we state how to break supersymmetry in heterotic string theories, then the rest will be similar to the previous two sections.

To break supersymmetry in heterotic string theories we use certain topological twists, the theory of topological twists is beyond the scope of this review for more details see [34].
In the upcoming discussion, we follow [35]:\\
Firstly, we divide the fermionic and bosonic variables into to two categories: the variables in the non supersymmetric theory and will be labeled by the subscript $A$ and variables not in the non supersymmetric theory which are labeled by the subscript $X$ if they belong to the $E_8 \times E_8$ theory, or $Y$ if they belong to the $SO(32)$ theory. This categorization is displayed in figure 1 in Ref [35].\\
Now we define the supersymmetry breaking twists as follows:\\
From $E_8 \times E_8$ theory we apply the twists
$$X^M_A \rightarrow X^M_A, \ \ \ \ \ \ \ X^M_X \rightarrow -X^M_X,$$
$$\chi_A^M \rightarrow -\chi_A^M, \ \ \ \ \ \ \ \chi_X^M \rightarrow \chi_X^M.$$
Here we assumed that all gauginos have positive chirality.\\
From $SO(32)$ theory we apply the twists
$$X^M_A \rightarrow X^M_A, \ \ \ \ \ \ \ X^M_Y \rightarrow -X^M_Y,$$
$$\chi_A^M \rightarrow -\chi_A^M, \ \ \ \ \ \ \ \chi_Y^M \rightarrow \chi_Y^M.$$
Here we assumed that all gauginos have negative chirality.

Applying these twists to the heterotic string action, we get a non supersymmetric string theory. Then we calculate the expectation value of the energy momentum tensor, equate the beta functions to zero, decompose the ambient manifold then deduce a condition on the internal manifold.
This procedure has a caveat that after the twist there are no preserved spinors on the internal manifolds. Thus, we can not derive any restrictive conclusion on it unless we consider manifolds with line bundles, this indeed gives the restriction we need to conclude that the manifold must be Calabi-Yau, but since the line bundles are considered as background flux (because they are on the manifold itself not derived from the theory), twists and flux compactifications will be discussed in more detail in chapter 7.\\

\includegraphics[width=\textwidth]{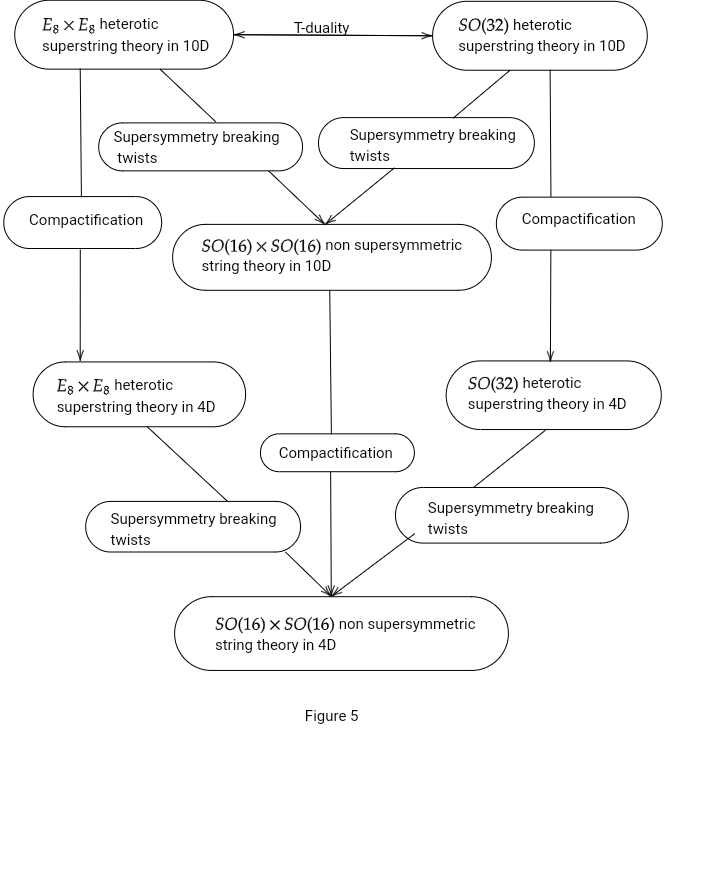}

\section{Generalized complex geometry}
Generalized complex geometry was introduced by Hitchin [36] and Gualtieri's PhD thesis [37]. The main motivation is to give a geometric meaning to the $B$-field in a rigorous way.
In this chapter we review the basic theory following [38].
\subsection{G-structures}
Firstly, we set up the notation used. Throughout this chapter let $M$ be a compact manifold with local patches $U_{\alpha}$, $E$ be a bundle over $M$ with fiber $F$ and section $s$, $TM$ be the tangent bundle over $M$, $T^*M$ be the cotangent bundle over $M$.

As discussed in chapter 2, the bundle $E$ is locally a product $M \times F$ but globally it can be a much more complicated structure. Therefore, we must define transformations to see how does the fiber transform from one patch to other neighbouring patches, in this way we may be able to examine global properties of the bundle. These transformations are called \textbf{transition functions}, and they play a key role in defining G-structures as we will see.
It is easy to see that the set of transition functions with the composition of functions for a group, this group is called \textbf{the structure group}.
The other key concept to define a G-structure is the tangent frame bundle.

 \textbf{Definition:} The tangent frame bundle denoted by $FM$ associated to $TM$ is a bundle over M whose fiber at each point $p \in M$ is the set of ordered basis of the tangent space of $M$ at this point $T_PM$.
 
Let's describe this bundle in a more practical way, consider two different patches $U_{\alpha}$ and $U_{\beta}$ which are nieghbourhoods of a point $p\in M$ with local trivializations ($p,\epsilon_{\alpha}$) and ($p,\epsilon_{\beta}$) respectively, where $\epsilon_{\alpha}$ and $\epsilon_{\beta}$ are the local basis for $T_p M$ on the patch $U_{\alpha}$($\beta$), physicists call them local frames of reference. Now suppose there is an overlap between the two patches, then the relations between the bases on the two patches are
$$\epsilon_{\alpha}=\epsilon_{\beta}t_{\alpha\beta},$$
where $t_{\alpha \beta}$ are the transition function from $U_{\alpha}$ to $U_{\beta}$.\\
But from differential geometry we know that the relations between basis on different patches are given by
$$\epsilon_{\alpha}=\frac{\partial x}{\partial \bar{x}}\epsilon_{\beta},$$
where $x=$($x_1,x_2,...,x_n$) such that $x_i$ are the normal coordinates over $U_{\alpha}$, and $\bar{x}$ is the same for $U_{\beta}$. Thus $\frac{\partial x}{\partial \bar{x}}$ is the Jacobian matrix that transforms one set of normal coordinated to the other.

Equating the last two equations we see that the group transition functions i.e. the structure group can be represented as a group of linear transformations i.e. the structure group is $GL(n,\mathbb{R})$. On the other hand the basis $\epsilon_{\alpha}$ can also be considered as a matrix (as each element for example $\epsilon_1$ is an element of the basis i.e. a vector). Thus, $ \epsilon_{\alpha}  \in GL(n,\mathbb{R})$. Then, collectively the fiber of the frame tangent bundle is $GL(n,\mathbb{R})$. This gives an equivalent definition for the tangent frame bundle.

\textbf{Equivalent definition:} A tangent frame bundle is the principal bundle whose fiber is the structure group.

We are interested in the cases where there exists an invariant structure over the manifold $M$, i.e. the same on all patches, this implies that the transition functions that change this invariant structure must be excluded leaving a subgroup G of the old structure group as the reduced structure group or equivalently, a reduced tangent frame bundle over $M$.

\textbf{Definition:}A manifold $M$ \textbf{has a G-structure} if it is possible to reduce its structure group to the group G, and the invariant used to do so is called the \textbf{G-invariant}.

\textbf{Example:} Suppose we have a globally defined linear transformation $$J: TM \rightarrow TM,$$ such that $$J^2=-Id,$$
i.e. an almost complex structure. In this case every matrix preserving this structure is a block matrix of two diagonal blocks each one is an $\frac{n}{2}\times \frac{n}{2}$ complex matrices. This implies that the group of all matrices preserving $J$ is isomorphic to $GL(\frac{n}{2},\mathbb{C})$ i.e. an almost complex manifold always has a $GL(\frac{n}{2},\mathbb{C})$ structure.

\subsection{Elements of generalized complex geometry}
The procedure is to replace the tangent bundle with the sum of the tangent and the cotangent bundle $TM \rightarrow TM \oplus T^*M$ called the generalized tangent bundle, and repeat everything in complex geometry with it.

We define a generalized tangent vector $\mathbb{X}$ as a sum of a tangent vector and a one form 
$$\mathbb{X}=X \oplus \xi,$$
where $X \in TM$ is called the vector part, and $\xi \in T^*M$ is called the one form part.\\
\subsubsection{The metric and the volume form}
On the generalized tangent bundle  we can define a metric in a natural way, unlike the tangent bundle, as follows: given two generalized vectors $\mathbb{X}=X \oplus \xi$ and $\mathbb{Y}= Y \oplus \eta$, we define the canonical metric by 
$$\mathcal{I}(\mathbb{X},\mathbb{Y})=\frac{1}{2}(\xi(Y)+\eta(X)).$$
It is easy to see that this metric has a signature of $(n,n)$, and is defined globally on the manifold thus, it reduces the structure group to $O(n,n)$.\\
Additionally, this metric comes with a natural volume form defined as
$$vol_{\mathcal{I}}=\frac{1}{(d!)^2}\epsilon^{i_1i_1,,,i_n}\partial_{x^{i_1}}\wedge...\wedge \partial_{x^{i_n}} \wedge \epsilon_{i_1i_1,,,i_n}dx^{i_1}\wedge...\wedge dx^{i_n},$$
where $\epsilon^{i_1...i_n}$ and $\epsilon_{i_1...i_n}$ are totally antisymmetric permutation symbols and $\partial_{x^{i_j}}=\frac{\partial}{\partial x^{i_j}}$.\\
Since there are two multiplied permutation symbols in the volume form therefore, the volume form does not depend on the orientation thus, the structure group is reduced once more to the transformations that do not change orientation i.e. $SO(n,n)$. This means that the generalized tangent bundle comes naturally with an $SO(n,n)$ structure on its base manifold.\\
The generators for $SO(n,n)$ are
$$
\begin{bmatrix}
A \ \ \ \ \ \ \ \  \  0 \\
\ \ 0 \ \ \ (A^{-1})^T
\end{bmatrix}, \ \ \ \ \ \ \ \ \
e^B=\begin{bmatrix}
Id \ \ \ 0\\
B \ \ \ Id
\end{bmatrix}, \ \ \ \ \ \ \ \ \
e^{\beta}=\begin{bmatrix}
Id \ \ \ \beta\\
0 \ \ \ Id
\end{bmatrix},
$$
where $A$ is the structure group of the tangent bundle, $B$ is a two form, $\beta$ is a two vector.\\
it is easy to see that the action of $e^B$ is
$$e^B(X \oplus \xi)= X+(\xi - B\cdot X),$$
where $\cdot$ is the interior product i.e. contraction. Similarly for $e^{\beta}$
$$e^{\beta}(X \oplus \xi)= (X-\xi \cdot \beta)+\xi.$$
The actions of $e^B$ and $e^{\beta}$ are usually called twists.\\

\subsubsection{(Twisted) Courant bracket}
The Courant bracket in generalized complex geometry is the counterpart of the Lie bracket in differential geometry, it is defined as follows

\textbf{Definition:} Let $\mathbb{X}=X \oplus \xi$ and $\mathbb{Y}=Y \oplus \eta$ be two generalized vector fields, their Courant bracket is given by
$$[\mathbb{X},\mathbb{Y}]_C=[X,Y]+\mathcal{L}_X \eta -\mathcal{L}_Y \xi -\frac{1}{2}d(\eta \cdot X- \xi \cdot Y),$$
where $[X,Y]$ is the Lie bracket of $X$ and $Y$, $\mathcal{L}_X$ is the Lie derivative with respect to $X$, $d$ is the exterior derivative and $\eta \cdot X$ is their contraction. 
This operation is well defined but it fails as a generalization of a Lie bracket because it is not a Lie bracket itself because it does not satisfy the Bianchi identity. To fix that we define the twisted Courant bracket by a closed three form $H$ as follows
$$[\mathbb{X},\mathbb{Y}]_H=[\mathbb{X},\mathbb{Y}]_C+(H \cdot X)\cdot Y.$$
It is called "twisted" because it satisfies
$$[e^B\mathbb{X},e^B\mathbb{Y}]_{H-dB}=e^B[\mathbb{X},\mathbb{Y}]_H.$$
This is indeed a Lie bracket unless $H=0$ it would be a Courant bracket.

Physically, we interpret $B$ as the B-field in string compactifications and $H$ to be its field strength i.e. a flux. Thus, if we turn on fluxes on compactifying we can use use generalized complex geometry to give a geometric meaning to the flux $H$ as the twist in the internal manifold and for the B-field as the action of the coordinated due to this twist, in the language of field theories $H$ is a physical field and $B$ is its potential.\\
\subsubsection{Polyforms and pure spinors}

\textbf{Definition:} A polyform $\phi$ is the formal sum of forms of different dimensions.\\
The interior product of polyforms and generalized tangent vectors is defined as 
$$\mathbb{X}\cdot \phi= X \cdot \phi+\xi \wedge \phi,$$
where $\mathbb{X}=X\oplus \xi$.\\
Now let's compute the action of an anticommutator of two generalized tangent vectors on a polyform, the result is
$$\{ \mathbb{X},\mathbb{Y}\}\cdot \phi=2\mathcal{I}(\mathbb{X},\mathbb{Y})\phi.$$
This is interesting because it means that generalized tangent vectors form a Clifford algebra (act as gamma matrices in quantum field theory language). Thus, polyforms transform in the spin representation of $Spin(n,n)$, we can use this to define different kinds of spinors like Majorana-Weyl spinors with a rigorous geometrical meaning but this is beyond the scope of this review.

Another interesting fact is that polyforms come with a natural bilinear form called \textbf{Mukai pairing} defined as follows:\\
Let $\phi_1$ and $\phi_2$ be two polyforms, their Mukai pairing is 
$$<\phi_1,\phi_2>=\phi_1 \wedge \sigma(\phi_2)|_{top},$$
where $\sigma$ is the operator reversing all indices of the polyform and $|_{top}$ denotes the projection of the top form i.e. the form with dimension n. This pairing is equivalent to the charge conjugation matrix in classical or quantum field theory.\\
Note that Mukai pairing is invariant under twisting i.e.
$$<e^B\phi_1,e^B\phi_2>=<\phi_1,\phi_2>.$$
So far we dealt with general polyforms and shown that they transform as spinors, now we define a special class of polyforms called pure spinors.\\
Firstly we need to define the null space of a polyform which is a similar to null spaces in linear algebra.

\textbf{Definition:} Let $\phi$ be a polyform, the \textbf{null space} of $\phi$ is the subbundle
$$L_{\phi}=\{ \mathbb{X} ; \mathbb{X}\cdot \phi=0 \},$$
i.e. all generalized tangent vectors annihilating $\phi$.

\textbf{Definition:} A subbundle $L$ is called isotropic if for every two generalized tangent vectors $\mathbb{X}$ and $\mathbb{Y}$ in L
$$\mathcal{I}(\mathbb{X},\mathbb{Y})=0.$$

\textbf{Definition:} We say that $L$ is maximally isotropic if its rank equals half the rank of the generalized tangent bundle.

\textbf{Definition:} A polyform $\phi$ is called a \textbf{pure spinor} if its null space is maximally isotropic.

A pure spinor and its isotropic null space induce a decomposition of the space of all polyforms as follows
$$\Lambda^* T^*M \otimes \mathbb{C}=\bigoplus_{-n/2 \leq k \leq n/2}U_k,$$
where $$U_k=\Lambda^{n/2-k}L\cdot \phi,$$
where we denote $\Lambda^* T^*M$ to be the space of all polyforms and $\Lambda^{n/2-k}$ to be the space of all $n/2-k$ dimensional sections of $L\cdot \phi$, this means that $U_k$ is what we get be applying an antisymmetric product of $n/2-k$ generalized tangent vectors of $L$ on $\phi$.
This decomposition is called a \textbf{filtration}.
\subsubsection{Twisted exterior derivative, generalized complex structures and integrability}
Now we define the analogue of the exterior derivative in generalized complex geometry i.e. the twisted exterior derivative.\\
Given a polyform $\phi$, its twisted exterior derivative is given by
$$d_H \phi=d\phi+H\wedge\phi,$$
which satisfies
$$d^2_H=0.$$

\textbf{Definition:} A pure spinor $\phi$ is called \textbf{a generalized Calabi Yau a la Hitchin} if
$$d_H\phi=0,$$
i.e. it is closed under twisted exterior derivative.\\
Now we introduce generalized complex structures.

\textbf{Definition:}A \textbf{generalized almost complex structure} is a map
$$\mathcal{J}:TM\oplus T^*M \rightarrow TM\oplus T^*M,$$
satisfying
$$\mathcal{J}^2=-Id,$$
and that the metric is Hermitian with respect to $\mathcal{J}$ i.e.
$$\mathcal{I}(\mathcal{J}\mathbb{X},\mathcal{J}\mathbb{Y})=\mathcal{I}(\mathbb{X},\mathbb{Y})$$
for every generalized tangent vectors $\mathbb{X}$ and $\mathbb{Y}$.\\
The previous two conditions reduce the structure group to $U(n/2,n/2)$. Note that we can define a generalized almost complex structure $\mathcal{J}$ from an almost complex structure $J$ as follows:
$$\mathcal{J}= \begin{bmatrix}
-J \ \ \ \ \ \  0\\
0 \ \ \ \ \ \ J^T
\end{bmatrix}.$$
This structure decomposes the bundle into two, each of which corresponds to an eigenvalue of $\mathcal{J}$ namely i and -i, the bundle corresponds to i is called $L_{\mathcal{J}}$ and the other is $\bar{L}_{\mathcal{J}}$.
The integrability of such structure can be studies using two different but equivalent approaches, the first straightforward one is analogous to the integrability of complex structures studied in chapter 2:

\textbf{Definition:} The generalized almost complex structure $\mathcal{J}$ is integrable if $L_{\mathcal{J}}$ is involutive under twisted Courant bracket i.e.
$$\mathbb{X},\mathbb{Y} \in \Gamma(L_{\mathcal{J}}) \implies [\mathbb{X},\mathbb{Y}]_H\in \Gamma(L_{\mathcal{J}}).$$
The other way is to firstly notice that we can associate a complex pure spinor with isotropic null space to every generalized almost complex structure, and this null space is the i eigenbundle of the generalized complex bundle. we denote this associated spinor by $\phi_{\mathcal{J}}$.

\textbf{Theorem:} An almost complex generalized bundle is integrable if and only if its associated spinor satisfies
$$d_H\phi_{\mathcal{J}}=\mathbb{X}\cdot \phi_{\mathcal{J}},$$
for some generalized tangent vector $\mathbb{X}$.

This can be used as an equivalent definition of integrability. As in the case of complex geometry, an integrable generalized almost complex bundle is called a generalized complex bundle.

\subsection{Generalized Calabi-Yau manifolds}
In this section the aim is to define the generalized Calabi-Yau geometry which is of utmost importance in flux compactifications as we will see the next chapter.

As we have seen in the previous chapter the existence of a globally defined spinor on the internal manifold i.e. some preserved supersymmetry, forces the geometry to be Calabi-Yau i.e. to have $SU(n/2)$ holonomy group on its patches. In the G-structure language we say that supersymmetry forces an $SU(n/2)$ structure on the internal manifold. Here we define the generalized version of this situation.
We begin by defining the $U(n/2) \times U(n/2)$ structure as follows:

\textbf{Definition:} A $U(n/2)\times U(n/2)$ structure consists of two generalized almost complex structures $\mathcal{J}_1$ and $\mathcal{J}_2$ such that they commute and the metric $-\mathcal{I}\mathcal{J}_1\mathcal{J}_2$ is positive definite. This structure is called \textbf{H twisted generalized Kahler structure} if the two generalized almost complex structures are H integrable.

\textbf{Theorem:} The generalized tangent bundle has a $U(n/2) \times U(n/2)$ structure if and only if there exists pure spinor bundles $\Psi_1$ and $\Psi_2$ such that they satisfy:
\begin{enumerate}
    \item $\Psi_2 \in \Gamma(U_0)$, where $U_i$ is a filtration associated to $\Psi_1$. Equivalently, $\Psi_1 \in \Gamma(V_0)$, where $V_i$ is a filtration associated to $\Psi_2$.
    \item The metric on the generalized complex structures to which $\Psi_1$ and $\Psi_2$ are associated is positive definite.
\end{enumerate}
The obstruction of introducing globally defined spinors on this structure is the overall factor ambiguity of $\psi_1$ and $\Psi_2$, to remove this ambiguity we reduce the structure group to $SU(n/2) \times SU(n/2)$ be imposing the normalization condition
$$<\Psi_1,\bar{\Psi_1}>=<\Psi_2,\bar{\Psi_2}>=0,$$
where $<.>$ is the Mukai pairing and $\bar{\Psi_i}$ is the complex conjugate of $\Psi_i$.
In this case $\Psi_1$ and $\Psi_2$ are globally defined, physically this means that there is some preserved supersymmetry in the resulting theory. We can now define the generalize Calabi-Yau geometry which will be used in the next chapter.

\textbf{Definition:} A \textbf{generalized Calabi-Yau geometry a la Gualtieri} is an $SU(n/2) \times SU(n/2)$ structure such that 
$$d_H \Psi_1=0,$$
and $$d_H\Psi_2=0.$$
This geometry is also called \textbf{twisted generalised Calabi-Yau} geometry.

This will be used to model the compactification manifold of string theory with flux, this is remarkable because it allows us to describe fields and fluxes in string theory in a purely geometric way as we will see in the next chapter.

\section{Flux compactifications}
In the previous chapters we saw that string compactifications have some problems like non broken supersymmetry and the fact that these types of compactifications give rise to moduli which can be interpreted as a long range force carriers. To deal with these problems, we have to turn on some fluxes to give potentials to the moduli in order to stabilize them. In this chapter we discuss some models of flux compactifications. Since this topic is very diverse with many models being proposed, we only discuss the most generic model in each string theory type and refer to other models.

\subsection{Fluxes and charges}
By a flux we mean a differential form defined on the manifold in question. Fluxes representing potentials introduced in the theory whose field strength satisfy a certain Bianchi identity as we will see. We present the different types of potential forms (i.e. fluxes) below.

To define a charge, consider an n-form flux $F_{\mu_1 \mu_2 ... \mu_n}$, this represents the field strength of an (n-1) potential. We define the charge of the mentioned flux by 
$$Q=\int_{\partial \Sigma}F_{\mu_1 \mu_2 ... \mu_{n-1} 0},$$
where $\Sigma$ is the manifold the flux is defined on, and $\partial \Sigma$ is its boundary defined when setting the nth coordinate to a value, without loss of generality we can set it to zero. 
\subsubsection{Dirac quantization rule}
In string theory charges due to a flux are defined on D branes are represented by currents through the strings attached to these branes. Thus, be charge conservation the strings must form a closed loop or the branes must be wrapped on closed cycles, this must be satisfied by all the flux in the compactification scheme to be viable, this condition leads to Dirac quantization condition

$$\int_{\Sigma_n}F_n=2\pi N,$$
where $F_n$ is an n-form flux, $\Sigma_n$ is an n cycle and $N$ is a natural number. For more details on different approaches to derive this condition see [39].

The most important consequence (of Dirac quantization condition is that it rules out some compactifications for example the compactification on $AdS\times S^5/Z_2$, to solve this issue we have to introduce orientifolds with fractional charges to satisfy the conditions, these type of compactifications are beyond the scope of this review, for more details and how orientifolds of different dimensions are introduced see [40].

\subsubsection{Types of fluxes}
In this section we present the types of fluxes derived from string theory action directly i.e. not introduced by hand on the manifold.

Remember that superstring theories have two sectors: Neveau Schwarz (NS) sector and Ramond (R) sector each of which has its own potential forms. We focus here on the massless section of both sectors, in NS massless sector there is the metric tensor, the Kalb Ramond field ($B$) and the dilaton ($\Phi$), the metric can not be considered as a flux because it is a property of the manifold so the fluxes here are $B$ and $\Phi$, we will discuss the dilaton flux later.\\
we define the field strength form of $B$ as follows
$$H=dB,$$
which satisfies the Bianchi identitiy $dH=0$. 

For the massless R sector we have different potentials according to the theory, here we use the democratic formulation [41]. In this sector (called RR sector because it is derived from R sector of left and right movers) we can define three types of field strength forms

\begin{enumerate}
    \item $F_n=dC_{n-1}+m \text{e}^B,$
\item $F_n^{(10)}=dC-H\wedge C+m \text{e}^B,$
\item $F^{(10)}=\sum_n F_n^{(10)},$
\end{enumerate}
where $C$ in both cases is the potential with the right rank, and m is the mass parameter $m=F_0$. The Bianchi identity of the flux are given by
\begin{enumerate}
    \item $dF_n=0,$
    \item $dF^{(10)}_n=0,$
    \item $dF^{(10)}=H \wedge F^{(10)},$
\end{enumerate}
with another condition imposed by the democratic formulation
$$F^{(10)}_n= (-1)^{[n/2]} \star F^{(10)}_{10-n}, $$
where $[n/2]$ is the integer part of $n/2$.

Finally, we address the dilaton which is a scalar field i.e. a zero form, in theories allowing zero forms flux in their RR sector, the dilaton is combined with the form to form one flux for example in type IIA superstring theory, there is an RR zero form $C_0$ then the total flux  will be $\tau= C_0 +ie^{\Phi}$. In theories who do not allow zero form potentials the dilaton will represent the topological term in the action i.e. the coupling constant for gravitational interactions as usual in string theory.

\subsection{Compactifications on twisted tori}

We saw in chapter 3 that toroidal compactifications results in $\mathcal{N}=8$ supesrymmetric low energy models which is too much supersymmetry for a realistic model and the fact that a torus can not induce chiral fermions. These two problems can be solved by introducing an additional geometric structure to the torus, namely a twist, which is equivalent to turning on a flux in the string model. In this section we briefly review the resulting low energy model and the challenges it faces.
\subsubsection{Twisted tori}
We begin by defining a twisted torus mathematically as follows: Let $M$ be a d dimensional manifold with coordinates $x^i$, we equip the manifold $M$ with a basis of non vanishing one form $\sigma_m$ i.e. a rank n tensor $T$ on $M$ can be written as $T=T_{\mu_1 \mu_2 ... \mu_n}\sigma^1 \sigma^2...\sigma^n$ where $T_{\mu_1 \mu_2 ... \mu_n}$ are the tensor's components and $\mu_i=1,2,...,d \ \forall i\in \{1,2,...,n\}$.
This manifold with the one form basis is called a \textbf{twisted d dimensional torus by a group $G$} if
\begin{enumerate}
\item The coordinated $x^i$ satisfy periodicity conditions: $x^i$ is identified to $x^i+2\pi R^i$ where $R^i$ is the radius of the torus in the i-th direction (as each direction is a circle as the case of ordinary torus).
\item the form basis satisfy the structure equation
$$d\sigma_m+\frac{1}{2}f^m_{np}\sigma^n \wedge \sigma^p=0,$$
where the quantities $f^m_{np}$ are the structure constant of a lie group $G$ representing the twisting as will be explained later, these quantities satisfy the Jacobi identity 
$$f^q_{[mn}f^t_{p]q}=0.$$
where $[mnq]$ is a commutator of indices.
\end{enumerate}
To describe twisted tori are we have to note that a torus has two classes of non contractible closed curves, and a vector field defined on the torus can be transformed along the curves in a well defined manner. The main idea of twisting is to transform a vector field defined on the torus by the action of the group $G$ wherever it is transformed along a non contractible closed curve, it is similar to holonomy but for finite (not infinitesimal) non contractible curves, a simple example is shown in figure 6 below. 

\begin{figure}[h]
    \centering
    \includegraphics{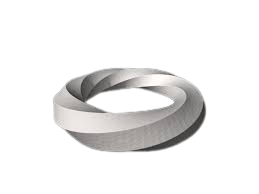}
    \caption{A simple example of a twisted torus when the twisting group $G$ is the rotation group, a vector field defined on this twisted torus will be rotated each time it revolves along a non contractible curve i.e. one rotation alone the curve will be equivalent to an action of the rotation group.}
    \label{fig:my_label}
\end{figure}

Note that when the twisting group $G$ is the trivial group, all the structure constants are zero and we recover the ordinary torus.\\
From this definition we can easily see that in the context of string theory, a twisted torus can be thought of as a torus with flux turned on [42], and from this we study the compactification on this geometry as flux compactification on a torus.

\subsubsection{The compactification model}
To perform a compactification of bosonic string theory on these type of tori, we use generalised geometry discussed in chapter 6 to get an $O(d,d)$ invariant theory where $d$ is the dimension of the twisted torus.

Firstly, we define the Lie algebra of symmetries of the theory as 
$$[Z_m,Z_n]=f^p_{mn}Z_p-K_{mnp}X^p,$$
$$[X^m,Z_n]=-f^m_{np}X^p,$$
$$[X^m,X^n]=0,$$
where $X^n$ are the generators of the gauge transformations of the flux, $Z_n$ are the generators of the diffeomorphisms on the twisted torus, $f^n_{mp}$ are the structure constants of the underlying Lie group of the twisted torus, and $K_{nmp}$ are constant coefficients satisfying the integrability conditions $K_{t[mn}f^t_{pq]}=0$.\\
To define a generalised geometry we define the generalised vector 
$$T_A=(Z_m,X^m),$$
where $A=1,2,...,2d$, which is an $O(d,d)$ vector as it satisfies 
$$[T_A,T_B]=f_{AB}^CT_C.$$
Secondly, we write the Lagrangian as
$$\mathcal{L}=R \star \mathbb{I}+\frac{1}{2}\star d\phi \wedge d\phi - \frac{\text{e}^{a\phi}}{2}\star F \wedge F,$$
where $R$ is the Ricci scalar of the manifold $M_{d+4}=\mathcal{M}_4\times \mathbb{T}^{d}$ such that $\mathbb{T}^{d}$ is the d dimensional twisted torus, $\mathbb{I}$ is the identity form, $\phi$ is the dilaton and $F=dB$ where $B$ is the Kalb Ramond field.\\
Now we separate the internal space i.e. the twisted torus from the internal Minkowski space and write the metric as
$$ds=g_{\mu\nu}dy^{\mu}dy^{\nu}+g_{mn}dx^mdx^n,$$
where $g_{\mu\nu}$ is the Minkowski metric with $\mu,\nu=1,2,3,4$, $y^{\mu}$ are the coordinates on the Minkowski space, $g_{mn}$ is the internal metric where $m,n=1,2,...,d$.

In this formalism we note that the external component of the Kalb Ramond field $B_{\mu\nu}$ transforms as a two form, the internal components $B_{mn}$ transforms also as a two form and $B_m$  transforms as a d vector additionally, there is another d vector $A^m$ representing the connection form on the torus. From these fields we can construct the generalised metric
$$\mathcal{M}^{AB}=
\begin{bmatrix}
g^{mn} & -B_{mn}g^{np} \\
-B_{mp}g^{np} & g_{mn}+g^{pq} B_{mp}B_{nq}
\end{bmatrix},
$$
along with the generalised vectors
$$\mathcal{A}^A= \begin{bmatrix}
A^m \\ B_m
\end{bmatrix},$$
representing the internal vector due to the connection on the generalised geometry and
$$\mathcal{F}^A=\begin{bmatrix}
F^m \\ G_m-B_{mn}F^n
\end{bmatrix},$$
where $G_m$ is defined in [46], this generalised vector representing the flux.

Using all these variables we can write the Lagrangian of the compactified theory as
$$\mathcal{L}=\text{e}^{\phi}(R\star \mathbb{I}+\star d\phi \wedge d\phi + \frac{1}{2}\star G_3 \wedge G_3 + \frac{1}{4}L_{AC}L_{BD}\star D\mathcal{M}^{AB} \wedge D\mathcal{M}^{CD}$$
$$-\frac{1}{2}L_{AC}L_{BD}\mathcal{M}^{AB} \star \mathcal{F}^C \wedge \mathcal{F}^D - \frac{1}{12}\mathcal{M}^{AD}\mathcal{M}^{BE}\mathcal{M}^{CF}t_{ABC}t_{DEF}$$
$$+ \frac{1}{4}\mathcal{M}^{AD}L^{BE}L^{CF}t_{ABC}t_{DEF}),$$
where $G_3$, $D\mathcal{M}^{AB}$ and $L^{AB}$ are defined in [45].

This models bosonic string theory on twisted tori, for supersymmetric counterparts see[43-47], the resulting theories contains chiral fermions solving the problem in toroidal compactifications without flux however, the supersymmetric models give $\mathcal{N}=4$ supersymmetric low energy models after adding flux which is an improvement compared to the $\mathcal{N}=8$ theories without flux, but still has too much supersymmetry to be a realistic model. For models with additional flux are introduced see [42].

\subsection{Compactifications of type II superstrings on Generalized Calabi-Yau manifolds}
In this section we review compactifications of types IIA and IIB superstring theories with flux on manifolds with $SU(3)$ structure, we focus on flux compactifications leaving $\mathcal{N}=1$ unbroken supersymmetry, and prove that any consistent such theory is equivalent to a compactification on generalised Calabi-Yau manifold reviewed in chapter 6.
\subsubsection{Torsion classes}
In this section we define torsion classes for manifolds with $\text{SU}(3)$ structure, as fluxes leading to $\mathcal{N}=1$ supersymmetric low energy effective models are completely specified by the manifold's torsion classes [48].\\
From basic differential geometry, we know that to any manifold we can define a torsion tensor. On manifolds with $\text{SU}(3)$ structure, the torsion tensor 
$T^m_{np}\in \Lambda^1 \otimes (\text{su}(3) \oplus \overline{\text{su}(3)})$
can be decomposed into tensors of different representations as follows
$$T^m_{np}=(\mathbf{1}\oplus \mathbf{1})\otimes (\mathbf{8}\oplus \mathbf{8}) \otimes (\mathbf{6}\oplus \mathbf{\bar{6}}) \otimes 2(\mathbf{3}\oplus \mathbf{\bar{3}}),$$
where the bar represents the adjoint representation. We call the first term i.e. $\mathbf{1}\oplus \mathbf{1}$ the first torsion class of the manifold, obviously it represents a complex scalar and is denoted by $W_1$. The second term is a complex $(1,1)$ form, it is called the second torsion class and is denoted by $W_2$. The third term is a real formal sum of forms $(2,1)+(1,2)$ and is denoted by $W_3$. The last term is in fact two copies of a representation so is two classes, one is a real one vector and one is a real one form, the former is called the fourth torsion class, denoted by $W_4$, and the later is called the fifth torsion class denoted by $W_5$.

For a manifold with both complex structure $J$ and a symplectic structure $\Omega$ we can write [49]
$$dJ=\frac{1}{2}\text{Im}(\bar{W_1}\Omega)+W_4 \wedge J+ W_3,$$
$$d\Omega = W_1J^2 + W_2 \wedge J + \bar{W_5} \wedge \Omega.$$
Having written the differentials of the complex and symplectic structures in terms of torsion classes, we can classify manifolds with $\text{SU}(3)$ structures only by knowing their torsion classes as in [49], this will be used in the next section to specify the types of manifolds used in the compactifications we are considering.

\subsubsection{Type II compactifications with flux}
Firstly, we answer the question: why do we consider only manifolds with $\text{SU}(3)$ structure? The answer is that we require that the manifold is Calabi-Yau if all the fluxes vanish to match the results in chapter 5. Thus, an $\text{SU}(n)$ structure is required and we take $n=3$ so the manifold will be 6 dimensional.\\

Now, we write the supersymmetry transformations and derive the $\mathcal{N}=1$ solutions.
For type II superstrings, the supersymmetry transformations are
$$\delta \psi_M=(\nabla_M+\frac{1}{8}H_{MNP}\Gamma^{NP}\Gamma_{11}+\frac{e^{\Phi}}{16}\sum_{n=0}^{9}\frac{1}{n!}F_{N_1N_2...N_n}\Gamma^{N_1N_2...N_n}\Gamma_M\Gamma_{11}^{[\frac{n}{2}]}\sigma^1)\epsilon,$$
$$\delta \lambda= (\Gamma^N\partial_N \phi + \frac{1}{2} \Gamma^{MPQ}H_{MPQ} \Gamma_{11} + \frac{1}{8}\text{e}^{\phi}\sum_{n=0}^9 \frac{(-1)^n}{n!}(5-n)F_{P_1 P_2...P_N}\Gamma^{P_1P_2...P_N} \Gamma_{11}^{[n/2]}\sigma^1)\epsilon $$
for type IIA and

$$\delta \psi_M=(\nabla_M - \frac{1}{8}H_{MNP}\Gamma^{NP}\sigma^3+\frac{e^{\Phi}}{16}[\sum_{n \  odd}\frac{1}{n!}F_{N_1N_2...N_n}\Gamma^{N_1N_2...N_n}\Gamma_M\sigma^1+$$ $$ \sum_{n \ even} \frac{i}{n!}F_{N_1N_2...N_n}\Gamma^{N_1N_2...N_n}\Gamma_M\sigma^2])\epsilon,$$

$$\delta \lambda= (\Gamma^N\partial_N \phi - \frac{1}{2} \Gamma^{MPQ}H_{MPQ} \sigma^1 + \frac{\text{e}^{\phi}}{8}[\sum_{n \ odd} \frac{(-1)^n}{n!}(5-n)F_{P_1 P_2...P_N}\Gamma^{P_1P_2...P_N} \sigma^1 + $$ $$i \sum_{n \ odd} \frac{(-1)^n}{n!}(5-n)F_{P_1 P_2...P_N}\Gamma^{P_1P_2...P_N} \sigma^2])\epsilon $$
for type IIB, where $\psi_M$ are the two gravitinos, $\lambda$ is a vector containing the two dilatinos, $\epsilon$ is the supersymmetry parameter i.e. supersymmetry spinor, $\Gamma$ are the Dirac matrices in different dimensions, $F$ are the RR fluxes, $H$ are the NS fluxes, $\phi$ is the dilaton, $\nabla$ is the covariant derivative, $\sigma^i$ are the Pauli matrices. 
Note that when all fluxes vanish we get the results in chapter 5.

As explained before, we can split the supersymmetry Weyl spinors into internal and external parts as follows
$$\epsilon^1=\xi^1_+ \otimes \eta_+ + \xi^1_- \otimes \eta_-,$$
$$\epsilon^2=\xi^2_+ \otimes \eta_- + \xi^2_- \otimes \eta_+,$$
for type IIA, and
$$\epsilon^1=\xi^1_+ \otimes \eta_+ + \xi^1_- \otimes \eta_-,$$
$$\epsilon^2=\xi^2_+ \otimes \eta_+ + \xi^2_- \otimes \eta_-,$$
for type IIB, where $\xi_{1,2}$ are the 4D spinors, $\eta^{1,2}$ are the 6D spinors and the subscripts $+$ and $-$ label the chirality of the spinors.

This setup in general gives $\mathcal{N}=2$ supersymmetry as discussed in chapter 5, here we can adjust the fluxes so that the two spinors are not independent i.e. we require the 4D spinors to be proportional then solve for the vacuum solutions using the adjusted flux, in practice we impose the condition that the 4D spinors are proportional then solve for the fluxes representing the required vacua.\\
The imposed conditions are

$$\xi^1=a\xi,$$
$$\xi^2=b \xi,$$
where $a$ and $b$ are complex functions. This means that the two 4D spinors are proportional to a common spinor and thus the resulting vacuum will preserve one globally defined spinor i.e. $\mathcal{N}=1$ supersymmetric.

Substituting in the supersymmetry transformations and equate to zero, we get a set of equations relating possible fluxes with the torsion classes presented in the previous section, after rather tedious calculations [50] we get the set of $\mathcal{N}=1$ solutions in terms of torsion classes as following tables taken from [48]
 \begin{figure}[h]
    \centering
     \includegraphics[scale=0.7]{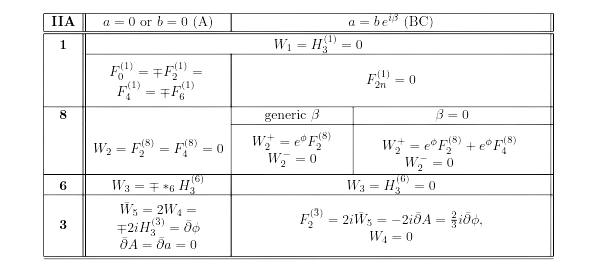}
     \caption{The set of all $\mathcal{N}$=1 solutions of type IIA superstring theory. The columns represent the solutions and the rows represent the fluxes in each representation.}
     \label{fig:my_label}
 \end{figure}

\begin{figure}[h]
    \centering
    \includegraphics[scale=0.7]{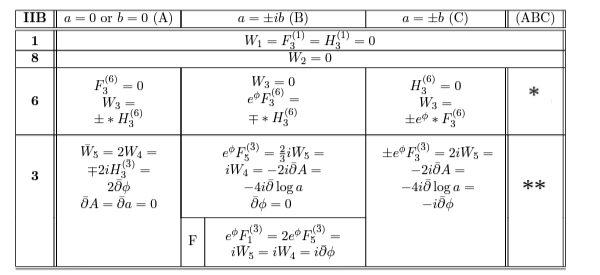}
    \caption{$\mathcal{N}$=1 solutions for type IIB superstring theory with the last column is displayed separately}
    \label{fig:my_label}
\end{figure}

\newpage

$$2abW_3=\text{e}^{\phi} (a^2+b^2) \star_6 F_3^{(6)},$$

\begin{equation}
(a^2-b^2)W_3=-(a^2+b^2) \star_6 H_3^{(6)}, \tag{*}
\end{equation}

$$2ab H_3^{(6)}=-\text{e}^{\phi} (a^2-b^2) F_3^{(6)},$$
where $\star_6$ is the 6 dimensional Hodge star operation and the superscript represents the representation the flux in,

$$\text{e}^{\phi} F_3^{(\bar{3})}=\frac{-4iab(a^2+b^2)}{a^4-2ia^3b+2iab^3+b^4} \bar{\partial}a,$$
$$\text{e}^{\phi} F_5^{(\bar{3})}=\frac{-4iab(a^2-b^2)}{a^4-2ia^3b+2iab^3+b^4} \bar{\partial}a,$$
$$H_3^{(\bar{3})}=\frac{-2i(a^2+b^2)(a^2-b^2)}{a^4-2ia^3b+2iab^3+b^4} \bar{\partial}a,$$

\begin{equation}
W_4=\frac{2(a^2-b^2)^2}{a^4-2ia^3b+2iab^3+b^4} \bar{\partial}a, \tag{**}
\end{equation}

$$\bar{W}_5=\frac{2(a^4-4a^2b^2+b^4)}{a^4-2ia^3b+2iab^3+b^4} \bar{\partial}a,$$
$$\bar{\partial}A=\frac{4(ab)^2}{a^4-2ia^3b+2iab^3+b^4} \bar{\partial}a,$$
$$\bar{\partial}\phi=\frac{2(a^2+b^2)^2}{a^4-2ia^3b+2iab^3+b^4} \bar{\partial}a.$$

These represent all possible $\mathcal{N}=1$ supersymmetric low energy models from type II superstring theory.

\subsubsection{Generalised geometry formulation}

In this section we derive all the previous results using generalised geometry reviewed in chapter 5, we prove that all the solutions can be encoded in 4 equations and are compactifications on twisted generalised Calabi-Yau manifolds.

Here we have a complex structure $J$ and a symplectic structure $\Omega$ on a generalised manifold with $\text{SU}(3)\oplus \text{SU}(3)$ structure. In this case we can define two no where vanishing bispinors [51,52]
$$\Phi_+=\eta_+ \otimes \eta_+^{\dagger},$$
$$\Phi_-=\eta_+ \otimes \eta_-^{\dagger}.$$
On the generalised manifold we can consider the two bispinors as two components of one generalised bispinor i.e. one spinor is conserved so the solutions have $\mathcal{N}=1$ supersymmetry.\\
Now we substitute in the supersymmetry transformations and equate to zero (the full calculations are done in [53]), the resulting equations are:
For type IIA
$$\text{e}^{-2A+\phi}(d+H\wedge)(\text{e}^{2A-\phi}{a\bar{b}\Phi_+})=0,$$
$$\text{e}^{-2A+\phi}(d+H\wedge)(\text{e}^{2A-\phi}{ab\Phi_-})=dA \wedge \bar{a}\bar{b}\bar{\Phi_-}-\frac{\text{e}^{\phi}}{16}[(|a|^2-|b|^2)F_{IIA+}-i(|a|^2+|b|^2) \star F_{IIA-}],$$
where $F_{IIA\pm}=F_0 \pm F_2 + F_4 \pm F_6$.\\ 
The first equation tells us that any compactification manifold must be twisted generalised Calabi-Yau and the second equation gives information about the type of twisted generalised Calabi-Yau used for each solution i.e. the equation used to tell us for each choice of flux which manifold must be used.\\
For type IIB
$$\text{e}^{-2A+\phi}(d-H\wedge)(\text{e}^{2A-\phi}{ab\Phi_-})=0,$$
$$\text{e}^{-2A+\phi}(d+H\wedge)(\text{e}^{2A-\phi}{ab\Phi_-})=dA \wedge \bar{a}b\bar{\Phi_+}+\frac{\text{e}^{\phi}}{16}[(|a|^2-|b|^2)F_{IIB+}-i(|a|^2+|b|^2) \star F_{IIB-}],$$
where $F_{IIB\pm}=F_1 \pm F_3 + F_5$.\\ 
The first equation says that the manifold is twisted generalised Calabi-Yau and the second gives the information about classifications of the solutions.\\

\subsection{Heterotic string theory compactifications with flux}
As shown in chapter 4, low energy models of heterotic string theories are already $\mathcal{N}=1$ supersymmetric but it is important to consider flux compactifications because the moduli problem in heterotic string theories are more severe than type II theories, this is due to the back reaction on the geometry when introducing new objects; in type II theories back reactions are always controlled and can be dealt with perturbatively, in heterotic string theories this is not always the case. Thus, we need to introduce flux to stabilize the moduli in a way that control all possible changes in the geometry.

In this section we review flux compactifications of a generic heterotic superstring theory with only $H$ flux turned on.\\
Following the analysis of [54], we begin by writing down the supersymmetry transformations and equating them to zero to get
$$(\nabla_M+\frac{1}{8}H_{MNP}\Gamma^{NP})\epsilon=0,$$
$$(\Gamma^M \partial_M \phi)\epsilon=0,$$
$$\Gamma^{MN}F_{MN}\epsilon=0,$$
the third equation is trivially satisfied in our case as only $H$ flux are on.\\
Decomposing $\epsilon$ into 4D and 6D parts
$$\epsilon = \xi \otimes \eta + \bar{\xi} \otimes \bar{\eta}, $$
where $\xi$ is the 4D part, $\eta$ is the 6D part, $\bar{\xi}=(B^{(4)}\xi)^*$ and $\bar{\eta} = (B^{(6)}\eta)^*$,
the first two equations imply the existence of a complex and symplectic forms as follows
$$J_m^n=-i\eta^{\dagger}\Gamma_m^n \eta,$$
$$\Omega_{mnp}=\bar{\eta}^{\dagger} \Gamma_{mnp}\eta.$$
Substituting in the equations, using the Bianchi identity for the flux, we get
$$H_{mnp}=3J_{[m}^r\nabla_|r|J_{np]},$$
where the commutator is over $m,n,p$ only and
$$H_{mnp}J^{np}=-4J_m^p\partial_p\phi.$$
The second equation tells us that the $H$ flux is not primitive unless the dilaton is a constant but in this case we must impose the condition $H=0$ to get a supergravity solution[55,56], if we require that the flux is non zero we must have a non constant dilaton so a non primitive flux. In this case we decompose the flux into primitive and non primitive parts as follows
$$H_{mnp}=H^P_{mnp}+H^{NP}_{mnp}=(H_{mnp}-\frac{3}{4}J_{[mn}H_{p]rs}J^{rs})+\frac{3}{4}J_{[mn}H_{p]rs}J^{rs},$$
applying the first supersymmetry transformation we get 
$$H^{NP}=-\star(d\phi \wedge J).$$
Substituting back into the transformations we get the two equations
$$H=\star \text{e}^{2\phi} d(\text{e}^{-2\phi}J),$$
$$d(\frac{1}{2}\text{e}^{-\phi}J \wedge \text{e}^{-\phi}J)=0.$$
Performing the conformal transformation $\bar{J}=\text{e}^{-\phi}J$, the second equation will be
$$d(\bar{J}\wedge \bar{J})=0.$$
This defines a geometry called the conformally balanced geometry. There is an important special case where $H^P=0$, the first condition gives $d\bar{J}=0$ which is a geometry related to the Kahler geometry by  a conformal transformation so called conformally Kahler geometry.

To summarize, If the flux has a non zero primitive part a consistent low energy model with non zero $H$ flux must be due to a compactification on a conformally balanced manifold, while if the primitive part is zero the compactification manifold must be conformally Kahler.

\subsection{Non-supersymmetric string theory compactifications with flux}
In this section we review flux compactifications of $\text{SO}(16)\times \text{SO}(16)$ heterotic non supersymmetric string theory, this theory is particularly important because it is the only non relativistic string theory with no tachyons [57].

It was shown that the low energy effective models of the $\text{SO}(16)\times \text{(SO)(16)}$ heterotic non supersymmetric string theory are identical to those of $E_8 \times E_8$ and $\text{SO}(32)$ heterotic supersymmetric string theories[36,58]. Thus compactifications of this theory can be deduced directly from supersymmetric heterotic theories compactifications i.e. if we want to compactify with certain flux, we can do the compactification on the supersymmetric counterpart then apply a supersymmetry breaking twist as explained in figure 5. 

\section{Applications to cosmology}
Having discussed string compactifications, now we present some applications to cosmology. In this chapter we discuss two important applications: the swampland project and the cosmological constant problem and the attempts to solve it by trying to find De Sitter vacua from low energy effective models of string theory.
\subsection{The swampland project}
\subsubsection{Introduction}
The swampland program was proposed by Vafa [59] to answer an important question in effective field theory: Can every effective field theory be embedded in a consistent high energy theory of quantum gravity? or in string theory, Does every low energy models come from some consistent string compactification?\\
The answer to this question is no. There are some constraints that tell us whether the low energy model is a result of a consistent string compactification (more generally, a low energy model of consistent quantum gravity theory), these constraints are called the \textbf{swampland conjectures}. Effective field theories satisfy the swampland conjectures i.e. can be lifted to a consistent quantum gravity theory are said to be \textbf{in the landscape}, otherwise the theory is said to be \textbf{in the swampland}.

The swampland conjectures are in fact very restrictive and may be broken for some theories at high energies. 
Thus, the swampland conjectures, also called the swampland constraints, form a cone as drawn in figure 9, that is because the constraints become stronger and more restrictive as the energy scale goes higher. For comprehensive review of the project see [60].
The next figure illustrates nicely the idea of the swampland project.

In the next section we review the swampland conjectures one by one and explain what are their impact on the physics of string compactifications as well as the mathematical requirements for the theory to be consistent.
\begin{figure}
    \centering
    \includegraphics[scale=0.5]{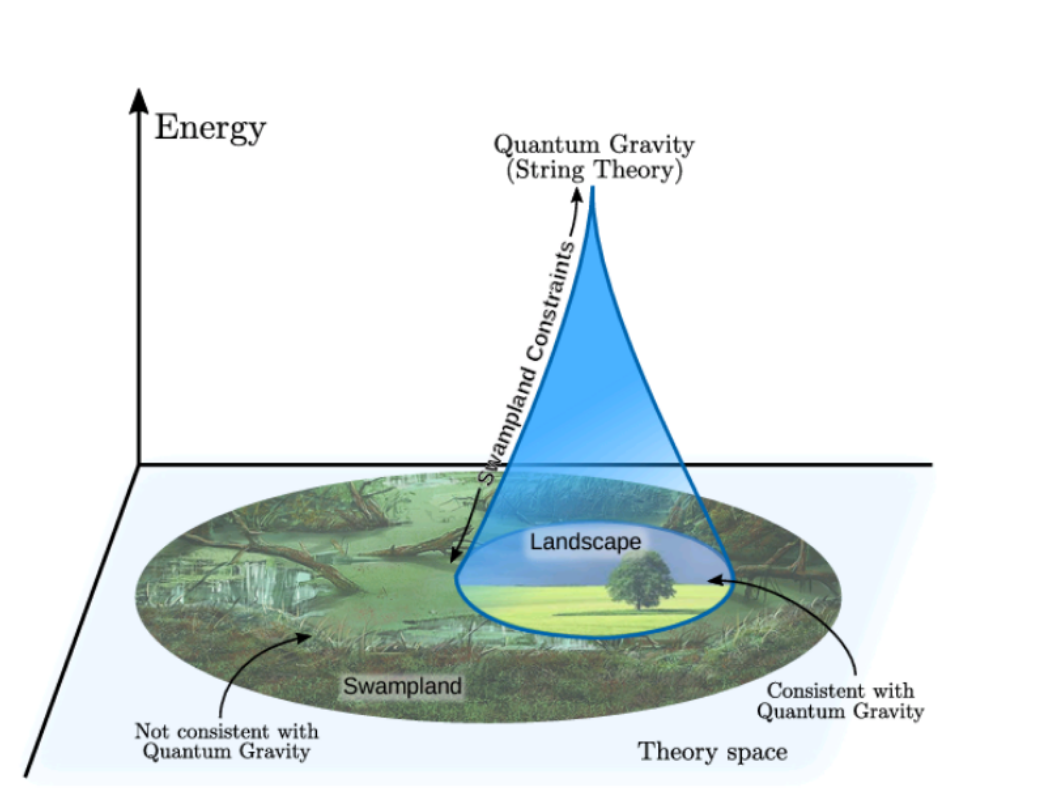}
    \caption{the horizontal plane represents the space of all low energy effective field theories, and the vertical axis representing the energy scale. For the theories in the swampland, the effective field theory description breaks down at certain energy scale and the high energy theory is not consistent with quantum gravity. For theories in the landscape many theories also break down at certain point if it breaks the swampland constraints but all consistent theories with quantum gravity must be in the landscape.}
    \label{fig:my_label}
\end{figure}

\subsubsection{Overview on the swampland conjectures}
The swampland conjectures can be divided into three groups displayed as rows in figure 10 below. 
\begin{enumerate}
    \item The first row i.e. the "no global symmetry" hypothesis, the cobordism conjecture and the completeness hypothesis, basically tell us that no global symmetries are allowed in a quantum gravity theory, gravity breaks all global symmetries.
    \item The second row i.e. the weak gravity conjecture and the distance conjecture give constraints on the spectrum of the low energy model and describe hoe the effective field theory breaks down for some theories at high energy scales.
    \item The third row i.e. the non supersymmetry AdS conjecture, the De Sitter conjecture and the AdS distance conjecture dive constraints on the possible consistent stable vacua allowed for the low energy effective model to be consistent with quantum gravity.
\end{enumerate}
\begin{figure}[h]
    \centering
    \includegraphics[scale=0.35]{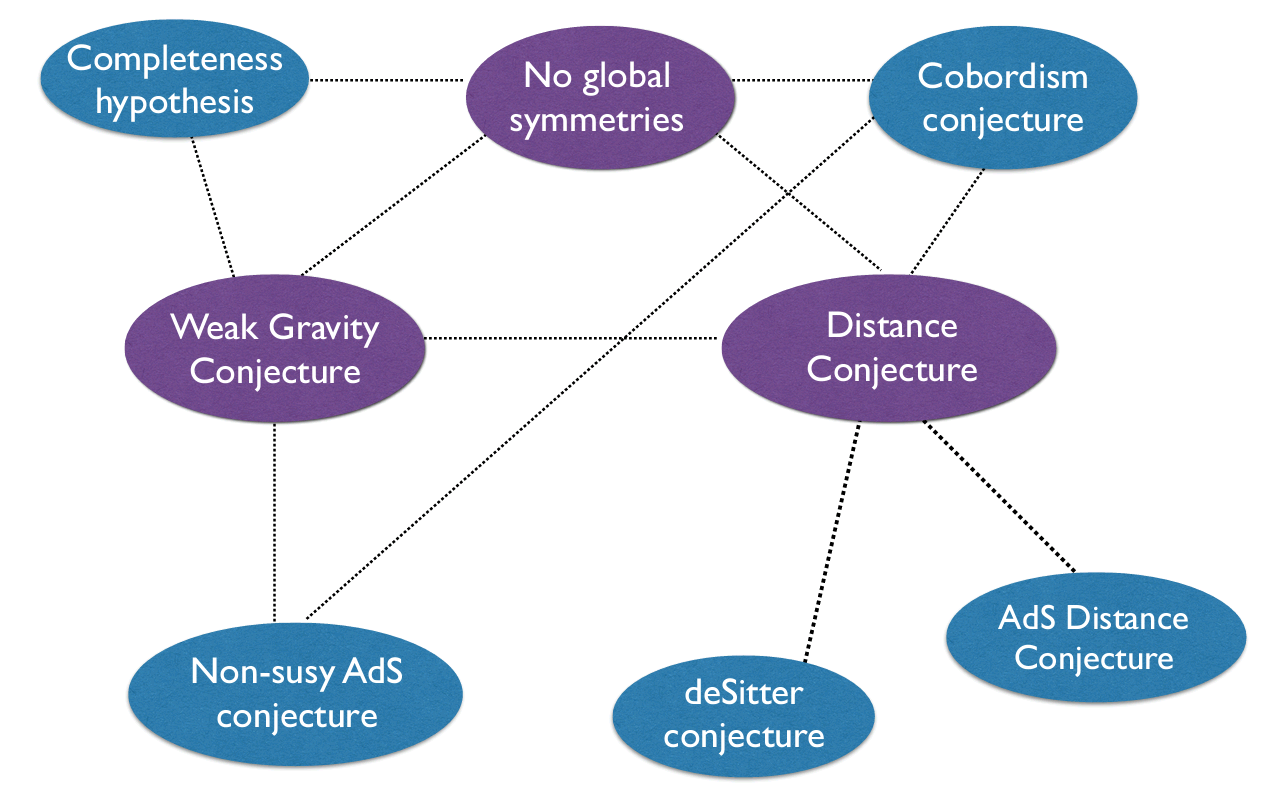}
    \caption{A schematic diagram showing the relations between the swampland conjectures.}
    \label{fig:my_label}
\end{figure}

Now we study the conjectures one by one.

\subsubsection{No global symmetries hypothesis}
The hypothesis [61-64] states that no global symmetries are allowed in any theory of quantum gravity.\\
Another equivalent statement: Any symmetry in a consistent quantum gravity theory must be either broken or gauged.

No global symmetries hypothesis is the most widely accepted among the swampland conjectures due to strong evidences from physics point of view especially cosmological arguments however,  there is no mathematical proof for it yet.\\
We motivate the hypothesis by considering the contrapositive argument i.e. what will go wrong if there are global symmetries in a theory of quantum gravity?
\begin{enumerate}
    \item The first motivation comes from black hole physics: Assume the existence of a global symmetry, then by Noether theorem we must have a global charge. Thus, a black hole can be charged under this symmetry. Upon decay the black hole will lose mass but not the global charge because Hawking radiation do not carry global charge, this will form a stable remnant which is not physical by itself. Moreover, since particles of any group representation carry the global charge, we can form a black hole easily as the remnant is very light, doing that we end up with an infinite number of stable black hole remnants which is of course non physical [61-65].
    \item Another motivation from black hole physics arguing that due to no hair theorem, a black hole can be completely specified by its mass, angular momentum and gauge symmetry charge. Thus, a global charge can not be measured i.e. has infinite uncertainty, this means that the black hole has infinite entropy and that is also non physical [66].
\end{enumerate}

Now we show that string theory satisfy this hypothesis as follows: It was proven that there can not be a global symmetry in the target space of a perturbative string theory even if we begin with a global symmetry in the parameter space [67].

Here we give an outline of the proof: Consider a global symmetry in the parameter space, the world sheet charge is given by
$$Q=\frac{1}{2 \pi i}\oint(j_zdz-j_{\bar{z}}d\bar{z}),$$
where $j_z$ and $j_{\bar{z}}$ are the currents of the global symmetry. The vertex operators for the currents are given by
$$j_z\bar{\partial} X^{\mu} \text{e}^{ikX},$$
$$\partial X^{\mu} j_{\bar{z}} \text{e}^{ikX}.$$
These operators correspond to massless gauge vector particles on the world sheet. Thus, a global symmetry in the parameter space is translated into a gauge symmetry on the world sheet which is the physical space according to string theory.

\subsubsection{Cobordism conjecture}
The conjecture states that in a consistent N dimensional quantum gravity theory compactified on a D dimensional internal manifold, all cobordism classes must vanish [68].
To motivate the conjecture, we ask what will go wrong if a cobordism class is non vanishing?\\
The answer is that we will get a global charge so a global symmetry. The global charge can be constructed as follows: Assume that we have a non vanishing cobordism class with cardinality M i.e. the compactification manifolds of different elements of the class are cobordant, then we have M higher dimensional manifolds each is the cobordism of an element of the class each of which is separate from the other completely, if we label these cobordism manifolds this label will be a global charge as no manifold from a certain class can change its class i.e. the label will be the same whatever interactions or fluxes are added.

An important implication to the cobordism conjecture is that it implies that every consistent compactification manifold of a quantum gravity theory is a boundary of some higher dimensional manifold. So we can think of the quantum gravity theory as this higher dimensional manifold, and every possible low energy effective field theory is a boundary of this manifold. This is a powerful statement because after compactification all possible effective field theories will be present but separated by domain walls.
\begin{figure}[h]
    \centering
    \includegraphics[scale=0.5]{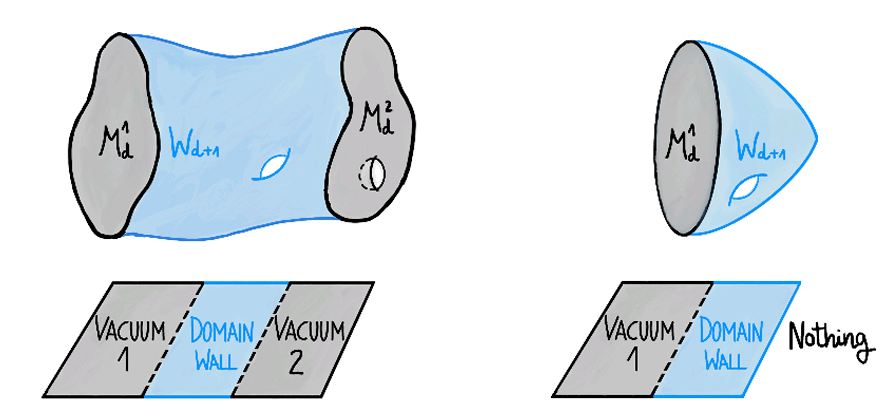}
    \caption{According to the cobordism conjecture every possible low energy model of a consistent quantum gravity theory is realised in the 4D space but separated by domain walls, in the case of a manifold with no boundary, the effective field theory space ends with the domain wall [69].}
    \label{fig:my_label}
\end{figure}

Before concluding the section we give two important notes:
\begin{enumerate}
    \item In mathematics literature, this topic is studied but using bordism not cobordism but the two approaches are exactly equivalent due to Yoneda lemma discussed in chapter 2(considering bordisms/cobordisms in the categorial sense). The two approaches are dual and can be transformed to each other at any point during any proof or argument, in this review we use cobordisms.
    \item In the case of flux compactifications where there are structures on the compactification manifolds, different types of cobordisms are defined and the cobordism conjecture apply for every type separately, in some cases we are required to add certain structures to some compactfications to satisfy the conjecture, this topic is beyond the scope of this review.
\end{enumerate}

\subsubsection{The completeness hypothesis}
The hypothesis states that for any low energy model coupled to gravity to be consistent with quantum gravity, these must exist physical states having all possible gauge charges consistent with Dirac quantization condition [70].

As usual we ask what can go wrong if this is not the case?\\
The answer is that we will get a global charge. Let the physical states have only a subset of the possible gauge charges allowed by Dirac quantization condition. Then, the left over charges form global charges of the theory. We illustrate this by an example: assume that the possible charges from Dirac quantization condition of some gauge symmetry is the set of all positive integers, but the physical states take only even charge values, then we can consider the gauge symmetry to have only the even charge spectrum and another symmetry with odd charge spectrum that acts globally i.e. we can label particles with odd charges and the original symmetry can not change this charge because interactions are present only within the even charge realm.

An important consequence of this conjecture is that it requires any continuous gauge group to be compact otherwise there would be infinite number of physical states to cover all the possible charges which is not pleasant in a physical theory.
This conjecture has some extensions to consider BPS states and supersymmetry but this is beyond the scope of the review.

\subsubsection{The weak gravity conjecture}
This conjecture has two versions [62,71]:
\begin{enumerate}
    \item The electric version: A gauge low energy theory coupled weakly to Einstein's gravity is consistent with quantum gravity if there exists an electrically charged (gauge charged in general) physical state such that its charge to mass ratio exceeds the charge to mass ratio of an extremal black hole (which by definition is of order one) i.e.
    $$\frac{Q}{M}\geq \frac{Q'}{M'}=O(1),$$
    where $Q$ and $M$ are the charge and mass of the physical state and $Q'$ and $M'$ are the charge and mass of an extremal black hole.
    \item The magnetic version: The cutoff at which a low energy model breakdown at is bounded from above by the gauge coupling 
    $$\Lambda \leq g M_p^{\frac{D-2}{2}},$$
    where $g$ is the gauge coupling constant, $M_p$ is the Planck mass and $D$ is the dimension of the compactification manifold.
\end{enumerate}
The electric version was proposed to allow the decay of extremal black holes without resulting in a naked singularity, the reasoning goes as follows: Consider an extremal black hole, the decay produce can increase or decrease the charge to mass ratio, if the product has a lower charge to mass ratio there is no problem as it will have a regular horizon and no naked singularity is produced, if the decay product has a higher charge to mass ratio it will have a naked singularity so it can not be a black hole thus, must be a particle i.e. a physical state with higher charge to mass ratio than that of an extremal black hole. Thus, we can see the conjecture as an escape route for black holes to have a particle to decay to.

A natural question is why do we need extremal black holes to decay at all? could we conjecture that extremal black holes do not decay?\\
The answer is no because in this case we will end up with infinite entropy with a similar argument as we presented in the global symmetry hypothesis.

The magnetic version was proposed to forbid the existence of global U(1) symmetry as it would be restored when the gauge coupling constant goes to zero. In this case according to the conjecture the cutoff will also goes to zero i.e. the theory would be in the swampland. For theories in the landscape the coupling constant can not go to zero so we can not restore the global U(1) symmetry.
This conjecture has a lot of evidence from string compactification spectra calculated throughout the review, the explicit calculations are shown in [72-74].

\subsubsection{The distance conjecture}
Before we state the conjecture we remind the reader that many low energy compactification models of string theory even with flux have large moduli spaces, which means that there is a large space of possible sets of parameters and this may not be controlled by the string compactification i.e. the theory can be physically and mathematically plausible but the moduli space has problematic points at which the theory is not physical.

In our case there may exist points in the moduli space at which there exists a global symmetry and the aim of the distance conjecture to make sure that these points do not cause problems in the low energy theory. We first define a metric on the moduli space, this is mathematically correct because of the functorical nature of the moduli i.e. the codomain of moduli functors are schemes as discussed in chapter 2, this allows us to define a metric and state that the metric is consistent throughout the compactification process because functors preserve structures (see chapter 2 for more details). We define the metric in a way that the problematic points with global symmetries are at infinite metric distance from a generic point in the moduli space. The distance conjecture assures us that the effective field theory description does not apply to these points so the low energy model is consistent with quantum gravity.
Having said that we state the distance conjecture: At infinite metric distance on the moduli space of a low energy effective field theory there exists  an infinite tower of states that becomes exponentially light [75].
The conjecture does the job as an infinite tower of light states breaks the effective field theory description.

In string compactifications the conjecture is called the emergent string conjecture: Any infinite metric distance limit in the moduli space of a string compactification theory is either a decompactification limit i.e. some compact dimensions become non compact or there exists a weakly coupled string becoming tensionless.

This version has the exact purpose of the general distance conjecture: In the problematic points of the moduli space we can not use effective field theory.

\subsubsection{AdS distance conjecture}
This conjecture is a case of the distance conjecture as flat spaces can be considered as limits of AdS spaces with the cosmological constant goes to zero. Considering the space of all AdS vacua and the metric to be such that the zero cosmological model point is at infinite distance, we get the conjecture:

Any AdS vacuum of a low energy effective model has an infinite tower of states that becomes light in limit $\Lambda \rightarrow 0,$ and the mass scales as $|\Lambda|^{\alpha}$ such that $\alpha$ is a real number [76].

\subsubsection{Non-supersymmetry AdS conjecture}
The conjecture states that there are no stable non supersymmetric vacua in a low energy model which is consistent with quantum gravity [77,78].
This means that supersymmetry is the only mechanism that prevents bubbles of nothing to form, the physics (and math) of bubbles of nothing is beyond the scope of this review, see [79,80] for more details.

We motivate this conjecture using the weak gravity conjecture in flux compactifications case as follows: Assume we have an AdS vacuum of a generic flux compactification theory with flux $F_n$ on. As discussed in chapter 7 flux are differential forms and can be thought of as field strengths of some potentials $F_n=dC_{n-1}$. Applying the weak gravity conjecture, we must have a state (in this case a brane of codimension 1) such that 
$$T\leq QM_P^2,$$
where $T$ is the tension of the brane and $Q$ is the electric charge. Since the space is non supersymmetric due to the BPS version of the weak gravity conjecture, the equality is ruled out and we are left with a strict inequality
$$T<QM^2_p.$$
A brane of codimension one with tension less than its electric charge is an instability in AdS space.

There is however a loophole in this conjecture is that it does not consider a general flux nor a generic theory of quantum gravity. Recently, a counter example was constructed out of type IIA massive supergravity [81]. Massive supergravity is not a new construction from string theory, in fact there are well known models of massive gravity from string compactifications on semi compact semi infinite surfaces [82]. Thus, the non supersymmetric AdS conjecture must be refined, the attempts to refine it are beyond the scope of the review but till date there is no final form of this conjecture.

\subsubsection{De Sitter conjecture}
The conjecture states that there are no stable or even metastable De Sitter vacua in any low energy model consistent with quantum gravity [83,84].

This conjecture has some evidence in specific string compactification models. Other versions were proposed to weaken it for example to say that apply only asymptotically. However, there are many attempts to find a De Sitter vacua from string compactifications as we will discuss in the next section.

\subsection{The possibility of finding De Sitter vacua}
Cosmological observations tell us that the universe is expanding i.e. has positive cosmological constant, in other words it has to be a De Sitter space. Along with some speculations that the conjecture may not be correct in this form, the conjecture was modified into what is called the \textbf{refined De Sitter conjecture} [85] : A low energy model can have metastable or rolling De Sitter vacua but not stable.

There are several refinement to the De Sitter conjecture but will not be discussed in this review [86-90].\\
The refined version allows us to construct cosmological models from string theory because a cosmological model does not need to be stable, it is physically viable to be metastable with decay time larger than the age of the universe, or a rolling model as some inflation models are.

We now describe some models realising metastable or rolling De Sitter vacua.
\subsubsection{Brane world cosmologies}
The main idea of brane world cosmologies is that the universe is assumed to be "printed" on a D brane in the higher dimensional theory. The interactions of open strings are restricted to D branes due to the boundary conditions of open strings however, closed strings motion are not restricted. For more detailed review on brane world cosmology see [91].

A subset of brane world cosmologies are the inflation inducing cosmologies where we assume there are two D branes one of them hosting the universe, the separation between the two branes represents a scalar field which drives inflation, this setup is shown in figure 12 taken from [91].
\begin{figure}[h]
    \centering
    \includegraphics[scale=0.7]{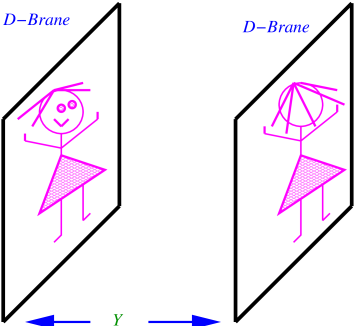}
    \caption{In inflation inducing brane world cosmologies there are two D branes, the universe is hosted by a D brane with the separation Y between them represents the inflaton which drives inflation.}
    \label{fig:my_label}
\end{figure}
One of the models realising De Sitter vacuum is the D3- anti D3 model where the universe is on a D3 brane and there is an anti D3 brane in the theory with distance Y between them [91], this model is an inflationary model as explained.

The De Sitter vacuum was constructed by turning on many flux that the tadpole cancellation condition require the existence of anti D3 branes, this introduces a potential which has a De Sitter metastable false vacuum, the decay time of this false vacuum is much larger than the age of the universe so the model is physically viable. For the detailed calculations see[92]. For other similar brane world cosmologies with metastable De Sitter vacua see [93-96].

\subsubsection{Warm inflation}
In the standard inflation scenario after the big bang there was an inflationary period where the universe expanded exponentially during which there were no particles, only a scalar field, called the inflaton, with large potential energy with low slope to support enough inflation. Particle are produced in the reheating phase where the slope of the potential becomes steep and the energy stored in the scalar field is converted into a hot soup of particles and antiparticles.

In warn inflation scenarios there is no separate reheating phase [97,98], the scenario assumes that particles are produced continuously during inflation i.e. the inflaton decays relatively slowly creating particles all the time not a steep fall like standard inflation.
In such models the potential of the inflaton can form De Sitter vacua, the condition on the potential to realise De Sitter vacua was derived in [99,100]. There are attempts to refine the De Sitter conjecture even more to incorporate Warm inflation models but this is still work under progress [101,102].

\subsubsection{Dynamical dark energy models}
The main issue with building cosmological models when accepting the De Sitter conjecture is that the universe is a De Sitter space. Dynamical dark matter models evade this problem by stating that the universe is in fact not the final vacuum state and is approaching the final AdS state i.e. it is only temporarily De Sitter [103-105].\\
These type of models have an important implication that dark energy is not constant as thought but is changing slowly towards an AdS state i.e. the cosmological constant is not a constant but a slowly varying function, hence the name of the model.
\begin{figure}[h]
    \centering
    \includegraphics[scale=0.5]{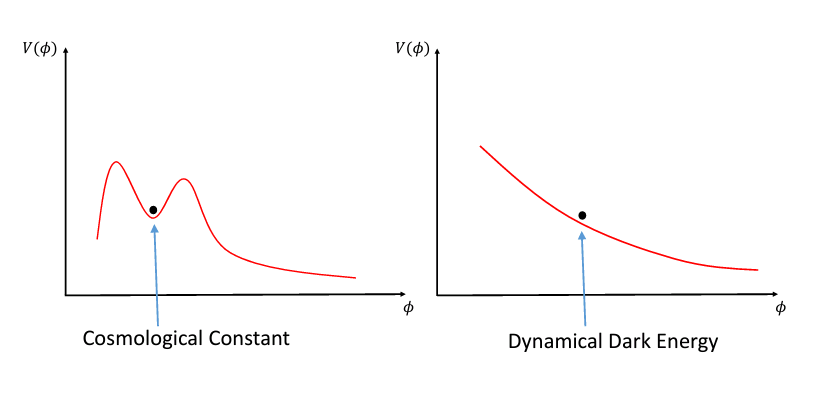}
    \caption{In the standard cosmological constant models the universe in thought to be in a (false) vacuum state as in the left graph. However, in dynamical dark energy models the universe is not a vacuum state at all, it is in its way to the final AdS vacuum. This evades the De Sitter conjecture as the De Sitter state of the universe is not a vacuum state.}
    \label{fig:my_label}
\end{figure}

\subsubsection{Other models}
There are other models aiming to either evade the De Sitter conjecture or modify it to be less restrictive about the De Sitter vacua allowed. Here we list some of these models.
\begin{enumerate}
    \item Multi field inflation where the inflation was driven by more than one inflaton field [106-108].
    \item Eternal inflation [109,110].
    \item Modified gravity models [111-113].
    \item String gas cosmology [114].
\end{enumerate}

\section{Non relativistic string theory}
In this chapter we will give a review on non relativistic string theory and construct a compactification model for non relativistic strings on a torus.

So far we have been discussing relativistic strings i.e. they give Einstein's gravity in their first order approximations. Non relativistic string theory on the other hand is designed to give Newtonian gravity instead, and used Newton Cartan geometry [115,116]. This theory like its relativistic counterpart is only consistent in 10 dimensions thus, as explained throughout the review, we have to compactify it to get 4D particle physics theory in this case coupled to Newton's gravity.

Non relativistic string theory was first studied by Gomis and Ooguri [117] where they considered closed non relativistic strings, further development has been made since then [118-123] but no where near its relativistic counterpart.

Till now there are no compactification models for non relativistic string theory, some aspects of the compactification were studied in [124] but a complete model has never been proposed.
Here we quickly review the basic theory then construct a compactification model on a torus.

We begin with the following sigma model action on the world sheet in curved spacetime
$$S=\frac{1}{4\pi \alpha'}\int d^2\sigma \sqrt{h}h^{\alpha \beta} \partial_{\alpha}X^{\mu} \partial_{\beta} X^{\nu} S_{\mu\nu}(X)-i\epsilon^{\alpha \beta} \partial_{\alpha}X^{\mu} \partial_{\beta} X^{\nu} A_{\mu\nu}(X)$$
$$+\sqrt{h}[\lambda \bar{D}X^{\mu}\tau_{\mu}(X)+\bar{\lambda}DX^{\mu}\bar{\tau}_{\mu}(X)+\lambda \bar{\lambda}U(X)+\alpha' R\Phi (X)],$$
where $\alpha, \beta =1,2$, $\mu,\nu=1,2,...,d$ such that d is the dimension of the ambient manifold, $\sigma^{\alpha}$ are the parameters of the world sheet, $d^2\sigma=d\sigma^1 d\sigma^2$, $h^{\alpha \beta}$ is the metric on the world sheet, $h=\det(h^{\alpha \beta})$, $S_{\mu \nu}$ is a symmetric tensor, $A_{\mu \nu}$ is an anti symmetric tensor, $\epsilon^{\alpha \beta}$ is the permutation symbol, $D$ and $\bar{D}$ are the covariant derivatives with respect to $X$ and $\bar{X}$ respectively, $\lambda$, $\bar{\lambda}$, $\tau$ and $\bar{\tau}$ are one forms on the world sheet, $U(X)$ is an unknown function, $R$ is the Ricci scalar of the world sheet and $\Phi$ is the dilaton.
This action is invariant under the Stuckelberg type symmetry (with the appropriate shift in the one forms in the action)
$$S_{\mu \nu} \rightarrow S_{\mu \nu}-2C_{(\mu}^{\alpha}\tau_{\nu)}^{\beta}\eta_{\alpha \beta},$$
$$A_{\mu \nu} \rightarrow A_{\mu \nu}-2C_{[\mu}^{\alpha}\tau_{\nu]}^{\beta}\epsilon_{\alpha \beta},$$
where $\eta_{\alpha \beta}$ in the Minkowski metric on the world sheet, the round brackets on indices means anti commutation and the square brackets means commutation and $C_{\mu}$ is an arbitrary matrix.

To fix this symmetry we introduce a longitudinal vielbeine $\tau_{\alpha}^{\mu}$ (by a longitudinal direction we mean along the string) such that $\tau_{\mu}^{\alpha}\tau^{\mu}_{\beta}= \delta^{\alpha}_{\beta}$, and impose the condition $\tau^{\mu}_{\alpha}S_{\mu\nu}=0$ i.e. $S_{\mu\nu}$ is fully transverse.
After fixing the symmetry the action can be written as
$$S=\frac{1}{4\pi \alpha'}\int d^2\sigma \sqrt{h}h^{\alpha \beta} \partial_{\alpha}X^{\mu}\partial_{\beta}X^{\nu}E_{\mu\nu}-i\epsilon^{\alpha \beta} \partial_{\alpha}X^{\mu}\partial_{\beta}X^{\nu}A_{\mu\nu}$$
$$+ \sqrt{h}(\lambda \bar{D}X^{\mu}\tau_{\mu}+\bar{\lambda}DX^{\mu}\bar{\tau}_{\mu}+\alpha'R\Phi).$$
In this process we defined longitudinal vielbeins, this motivated the definition of a transverse veilbeine $E_{\mu}^{A}$ where $A$ runs over the $d-2$ transverse directions to the string. The condition is then
$$S_{\mu\nu}=E_{\mu\nu}=\delta_{AB}E^A_{\mu}E^B_{\nu},$$
where the first equality is just a renaming.\\
The relations between longitudinal and transverse vielbeine are given by
$$\tau_{\mu}^{\alpha}\tau^{\mu}_{\beta}=\delta^{\alpha}_{\beta},$$
$$\tau_{\mu}^{\alpha}E^{\mu}_{A}=E_{\mu}^{A}\tau_{\alpha}^{\mu}=0,$$
$$E_{\mu}^AE^{\mu}_B=\delta^{A}_B,$$
$$\tau_{\mu}^{\alpha}\tau^{\nu}_{\alpha}+E_{\mu}^AE^{\nu}_A=\delta ^{\nu}_{\mu}.$$
This geometry is called \textbf{torsional string Newton Cartan geometry}, and is the geometry used in the theory of non relativistic strings.

The action on this geometry has a symmetries described by an algebra called \textbf{F-string Galilei algebra} whose generators are: $P_{A},P_{\alpha}$ generating longitudinal and transverse translations respectively, $M_{\alpha \beta} ,M_{AB}$ generating rotations in transverse and longitudinal directions respectively, $G_{A\alpha}$ generating Galilei boosts, $Q_{A},Q_{\alpha}$ are two generators due to the gauge nature of $A_{\mu\nu}$, and two new generators to close the algebra defined as
$$H_{\alpha} =c(P_{\alpha}+Q_{\beta}\tau^{\beta}_{\alpha}),$$
$$N_{\alpha} = \frac{1}{2c}(\epsilon_{\alpha}^{\beta}P_{\beta} +Q_{\alpha}).$$
The algebra has 10 non zero commutation relations defined in [116].\\

The theory is only consistent in (9+1)D i.e, there are 8 transverse space dimensions, we need to compactify the theory such that we have one transverse direction so that we end up with a 4D theory.

That was a highlight of the general theory. Now we present the compactification model on a torus. For simplicity, we consider flat specetime. To do that we simplify the action by integrating out all the one forms and setting the dilaton to zero. This gives the $\beta \gamma$ system action in the original paper by Gomis and Ooguri 
$$ S= \int \frac{d^2z}{2\pi} (\beta \bar{\partial}\gamma + \bar{\beta}\partial \bar{\gamma} + \frac{1}{4 \alpha_{eff}} \partial \gamma \bar{\partial} \bar{\gamma} +\frac{1}{\alpha_{eff}} \partial X^{\mu} \bar{\partial} X_{\mu}). $$
where $\gamma= X^1+X^2$, $\bar{\gamma}=X^1-X^0$, $z, \bar{z}$ are the complex coordinates on the worldsheet and $\alpha_{eff}$ is the finite effective string scale of non relativistic closed string theory, here it is equal to $\alpha'$ but we write it this way to match with the literature.

Knowing that the conformal dimensions of $\beta$ and $\gamma$ are $(1,0)$ and $(0,0)$ respectively, we deduce that $\gamma$ must admit winding and on a torus its mode expansion is given by
$$\gamma(z)=iE^{\mu}_r(w_{\mu}-B^{\mu}_m w_{\mu})R^m \log(z)+\sum_{i=-\infty}^{\infty}i \gamma_i z^{-i},$$
where $w_{\mu}$ is the winding number in the $\mu$-th direction, $B$ is the background field and $\gamma_i$ are the coefficients of expansion of the non compact directions.\\

From this we can calculate the Virasoro generators
$$L_n=-i\beta_n E^{\mu}_r(w_{\mu}-B^{\mu}_m w_{\mu})R^m + \sum_{m=-\infty}^{\infty} m \beta_{n-m} \gamma_m.$$
The solutions for the Virasoro constraint $L_0+\bar{L_0}=1$ is then
$$i \beta_0 E^{\mu}_r(w_{\mu}-B^{\mu}_m w_{\mu})R^m= -1 +\frac{\alpha_{eff}k^2}{4},$$
where $k$ is the momentum of the string along the non compact directions, the factor of $-1$ represents the conformal weight of the bosonic string, if we include NS sector strings, we replace it with the total conformal weight $N$, so the general solution is
$$i \beta_0 E^{\mu}_r(w_{\mu}-B^{\mu}_m w_{\mu})R^m= N +\frac{\alpha_{eff}k^2}{4},$$
and 
$$i \bar{\beta}_0 E^{\mu}_r(w_{\mu}-B^{\mu}_m w_{\mu})R^m= \bar{N} +\frac{\alpha_{eff}k^2}{4}.$$
We can then derive the closed string spectrum to be
$$ E^{\mu}_r(w_{\mu}-B^{\mu}_m w_{\mu})R^m (p_0-\sum_{i=1}^d \frac{n_i}{R^i})=\frac{1}{2\alpha_{eff}} (E^{\mu}_r(w_{\mu}-B^{\mu}_m w_{\mu})R^m)^2+ \frac{\alpha_{eff}k^2}{2}+ N + \bar{N},$$
where $d$ is the number of compact dimensions and $n_i$ are non negative integers.\\
Before analysing this spectrum, we note that the calculations for open string spectrum would be similar to the calculations in Gomis and Ooguri's paper and we deduce the same result that the open string spectrum is the same as in relativistic string theories.

Finally, we give some comments on the closed strings spectrum we deduced. We notice that the spectrum has tachyons if we considered closed strings only but are removed when considering superstrings which is the standard for string theory. The other important property is that there is no massless states in the closed string spectrum i.e. there is no graviton, and gravity is an instantaneous force. This was expected because non relativistic string theory is supposed to reduce to Newtonian gravity in which gravity is instantaneous. Thus, we conclude that the spectrum of non relativistic string theory is the same as the relativistic one except that the closed string spectrum is non relativistic and with no graviton. We also can calculate vertex operators and amplitudes using the same procedure as the original paper by Gomis and Ooguri using the Virasoro generators we calculated, we did not present it here as it is systematic and does not serve the purpose of the review.

Having constructed a compactification model for non relativistic strings, now we present possible directions for future work. We constructed a toroidal compactification model which is the simplest type of models, further more sophisticated models including Calabi-Yau compactifications would be interesting to investigate. Other questions involve the algebraic structures that can be added to non relativistic sigma models and it would be interesting to study the moduli spaces of such theories.

\section{Open problems}
In this chapter we present some open problems in string theory and the swampland project.
\subsection{Moduli stabilization problem}
Since we did not discuss moduli in this review, we briefly present this problem and refer to some attempts made to tackle it.

As explained on building compactifications models, we get massless scalars with no potentials and these are problematic from physical point of view. Thus we have to induce potentials to fix their vacuum expectation values i.e. stabilize the moduli. On turning on flux some moduli can be stabilized, in some models all moduli are stabilized, but a systematic method to stabilize moduli in a generic model is not yet known. The literature in this topic is extremely vast, for reviews on moduli stabilization see [125-127].

\subsection{The swampland project}
The swampland project is still under construction so there are many open problems within the project [128] and still many things to explore in this area. Here we list the most interesting five of the open problems(other than the problem of finding De Sitter vacua as we discussed it earlier):
\begin{enumerate}
    \item The weak gravity conjecture protects the theory from the problematic points in the moduli space as discussed before, the inverse question is not yet clear i.e. if we have a point of infinite metric distance in the moduli space (whether it is problematic or not), does it corresponds to a weak coupling limit?
    \item Mathematically, the weak gravity conjecture is defined only for continuous gauge symmetries but it is not yet clear how to define a counterpart for discrete gauge symmetries.
    \item In axion models with low codimension objects there are no extremal black holes. Thus the main motivation of the weak gravity conjecture does not exist anymore, so does the conjecture hold in these models?
    \item In the distance conjecture we appear to invent a metric to get rid of the problematic points and it was a successful approach, the question is: is there a coupling constant that goes to zero at infinite distance limit? if so this will further justify this approach.
    \item The distance conjecture is not yet fully understood in AdS/CFT models.
\end{enumerate}

\section{Conclusion}
In this review we studied string compactifications and how to get low energy 4 dimensional particle physics theories from high energy string theory, as well as some applications to cosmology.

We studied various string compactifications on non Calabi-Yau manifolds and found that although they give interesting spectra, they can not be realistic models and should be considered as toy models. For example low energy models from toroidal compactifications can not host chiral fermions, and compactifications to $AdS_5$ gives 4 dimensional theories on $AdS$ space which is different from what we observe i.e. the universe is a De Sitter space. Moreover, the models have too much supersymmetry.

We also studied compactifications on Calabi-Yau manifolds which was proven to give the right amount of supersymmetry in case of heterotic string theory, and $\mathcal{N}=2$ in case of type II string theories. We also reviewed compactifications of non supersymmetric string theory on Calabi-Yau manifold and found that it can be derived from heterotic string compactifications with an extra topological twist. These models however uses Calabi-Yau manifolds which give rise to many moduli, since moduli can mediate long range forces, these models can not be realistic models as they are, we must introduce additional structures i.e. flux to stabilize the moduli.

We then proceeded to study compactifications with flux i.e. flux compactifications, we found that flux can break supersymmetry and stabilize moduli. We studied toroidal compactifications with flux, which is equivalent to compactifications on twisted tori, this gives $\mathcal{N}=4$ supersymmetric low energy models, still not realistic but better than the $\mathcal{N}=8$ models without flux, we also found that twisted tori can host chiral fermions. we reviewed how to get $\mathcal{N}=1$ supersymmetric models from type II superstring theories and proved that the compactification manifold must be generalised Calabi-Yau. In the case of heterotic superstring theory, we found that on turning on $H$ flux, the compactification manifold must be conformally balanced or conformally Kahler depending on whether the flux has primitive part or no. Additionally, we studied non supersymmetric string theory compactifications with flux and found that the low energy models derived from it can be derived from heterotic superstring theory with similar flux with the extra topological twist we discussed earlier.

 We then presented some applications to cosmology. We introduces the swampland project and the swampland conjectures which give strict constraints on low energy models to be consistent with quantum gravity, we studied these constraints one by one and gave hints on various modifications to them. We found that the constraints are not yet in their final form, and the swampland project is still under construction which is an active research area at the moment. Then, we introduced some cosmological models motivated by the swampland project and string theory. The presented models satisfy the swampland conjectures but come with their own challenges, these models are also under construction.

 Then, we studied non relativistic string theory compactifications. We reviewed the basic theory and the algebra of the theory i.e. F string Galilei algebra, and its generators. Compactification models of non relativistic string theory were never been formulated, we present a simple compactification model and further research topics in the field.

Finally, we presented some open problems in the field including the problem of moduli stabilization. Although the main motivation of flux compactifications was to stabilize moduli, in most models not all the moduli were being stabilized. A systematic way to stabilize moduli in a generic string compactification model was being proposed but still under construction. We also listed some open problems in the swampland program, as explained the project is still under construction thus, there are many open problems and research ideas in this field and a lot of work to be done for the project to reach its final form.

\section{References}
\begin{enumerate}[label={[\arabic*]}]
\item Eilenberg, S., and Maclane, S. (1945). General Theory of natural equivalences. Transactions of the American Mathematical Society, 58(2), 231-294. 
\item Mac Lane, S. 1978. Categories for the working mathematician 2nd ed. New York: Springer-Verlag. 
\item Gross, M., Huybrechts, D., $\&$ Joyce, D. (2003). Calabi-Yau manifolds and related geometries. Berlin: Springer.
\item Nakahara, M. Geometry, Topology and Physics.
\item Hartshorne, R. (2010). Algebraic geometry. New York: Springer.
\item Shafarevich, I. (2014). Basic algebraic geometry 1. Berlin : Springer-Verlag.
\item West, P. (2012). Introduction to strings and branes. Cambridge: Cambrigde University Press.
\item Ibanez, L.$\&$ Uranga, A. (2012). String theory and particle physics. Cambridge: Cambridge University Press.
\item Fraiman, B., Graña, M., $\&$ Nuñez, C. (2018). A new twist on heterotic string compactifications. Journal Of High Energy Physics, 2018(9). doi: 10.1007/jhep09(2018)078.
\item Ginsparg, P. (1987). On toroidal compactification of heterotic superstrings. Physical Review D, 35(2), 648-654. doi: 10.1103/physrevd.35.648.

\item R. R. Metsaev and A. A. Tseytlin, “Type IIB superstring action in$ AdS_5 \times S^
5$
background”, Nucl. Phys. B533, 109 (1998), hep-th/9805028.
\item J. Polchinski, “Dirichlet Branes and Ramond-Ramond charges”,
Phys.Rev.Lett. 75, 4724 (1995), hep-th/9510017.
\item Green, M., Schwarz, J., $\&$ Brink, L. (1982). N = 4 Yang-Mills and N = 8 supergravity as limits of string theories. Nuclear Physics B, 198(3), 474-492. doi: 10.1016/0550-3213(82)90336-4.
\item Furutsu, H., $\&$ Nishiyama, K. (1991). Classification of irreducible super-unitary representations of$su(p,q/n)$. Communications In Mathematical Physics, 141(3), 475-502. doi: 10.1007/bf02102811.
\item Gunaydin, M., $\&$ Marcus, N. (1985). The spectrum of the S 5 compactification of the chiral N=2, D=10 supergravity and the unitary supermultiplets of U(2,2/4). Classical And Quantum Gravity, 2(2), L11-L17. doi: 10.1088/0264-9381/2/2/001.
\item Günaydin, M. (1998). Unitary supermultiplets of and M-theory. Nuclear Physics B, 528(1-2), 432-450. doi: 10.1016/s0550-3213(98)00393-9.
\item Schwarz, J. (2020). The $AdS_5 \times S^5$ superstring. Proceedings Of The Royal Society A: Mathematical, Physical And Engineering Sciences, 476(2240), 20200305. doi: 10.1098/rspa.2020.0305.
\item Bergshoeff, E., Kallosh, R., Ortín, T., $\&$ Papadopoulos, G. (1997). κ-symmetry, supersymmetry and intersecting branes. Nuclear Physics B, 502(1-2), 149-169. doi: 10.1016/s0550-3213(97)00470-7.
\item Duff, M., Lü, H., $\&$ Pope, C. (1998). AdS5 × S5 untwisted. Nuclear Physics B, 532(1-2), 181-209. doi: 10.1016/s0550-3213(98)00464-7.
\item Wingerter, A. (2022). Target Space Duality in String Theory (Phd). Rheinischen Friedrichs-Wilhelms-Universitat Bonn.
\item Giveon, A., Porrati, M., $\&$ Rabinovici, E. (1994). Target space duality in string theory. Physics Reports, 244(2-3), 77-202. doi: 10.1016/0370-1573(94)90070-1.
\item Heterotic String Theory on Anti-de-Sitter Spaces arXiv:hep-th/9805099v1.
\item Choi, K., 2005. Spectra of heterotic strings on orbifolds. Nuclear Physics B, 708(1-3), pp.194-214.
\item Dixon, L., Harvey, J., Vafa, C. and Witten, E., 1985. Strings on orbifolds. Nuclear Physics B, 261, pp.678-686.
\item Dixon, L., Harvey, J., Vafa, C. and Witten, E., 1986. Strings on orbifolds (II). Nuclear Physics B, 274(2), pp.285-314.
\item Candelas, P., Horowitz, G., Strominger, A., $\&$ Witten, E. (1985). Vacuum configurations for superstrings. Nuclear Physics B, 258, 46-74. doi: 10.1016/0550-3213(85)90602-9.
\item Graña, M. $\&$ amp; Triendl, H.M., 2017. String theory compactifications, Cham: Springer International Publishing. 
\item Hübsch, T., 1987. Calabi-Yau manifolds — motivations and constructions. Communications in Mathematical Physics, 108(2), pp.291–318. 
\item Polyakov, A.M., 1981. Quantum geometry of Bosonic Strings. Physics Letters B, 103(3), pp.207–210. 
\item Friedan, D., Martinec, E. $\&$ amp; Shenker, S., 1986. Conformal invariance, supersymmetry and string theory. Nuclear Physics B, 271(1), pp.93–165. 
\item Callan, C.G. et al., 1985. Strings in background fields. Nuclear Physics B, 262(4), pp.593–609.  
\item Rogers, F.A., 2007. Supermanifolds: Theory and applications, Singapore: World Scientific. 
\item Martinec, E.J., Robbins, D. $\&$ amp; Sethi, S., 2011. Non-supersymmetric string theory. Journal of High Energy Physics, 2011(10). 
\item Calaque, D. $\&$amp; Strobl, T., 2015. Mathematical aspects of quantum field theories, Cham: Springer. 
\item Blaszczyk, M. et al., 2015. Calabi-Yau compactifications of non-supersymmetric heterotic string theory. Journal of High Energy Physics, 2015(10). 
\item Hitchin, N., 2003. Generalized calabi-yau manifolds. The Quarterly Journal of Mathematics, 54(3), pp.281–308. 
\item Gualtieri, M. $\&$ Hitchin, N.J., Generalized complex geometry. thesis. 
\item Koerber, P., 2010. Lectures on generalized complex geometry for physicists. Fortschritte der Physik, 59(3-4), pp.169–242. 
\item Heras, R., 2018. Dirac quantisation condition: a comprehensive review. Contemporary Physics, 59(4), pp.331-355.
\item Tachikawa, Y. and Yonekura, K., 2019. Why are fractional charges of orientifolds compatible with Dirac quantization?. SciPost Physics, 7(5).
\item Bergshoeff, E., Kallosh, R., Ortín, T., Roest, D. and Proeyen, A., 2001. New formulations of D = 10 supersymmetry and D8-O8 domain walls. Classical and Quantum Gravity, 18(17), pp.3359-3382.
\item Hull, C. and Reid-Edwards, R., 2009. Flux compactifications of string theory on twisted tori. Fortschritte der Physik, 57(9), pp.862-894.
\item Dall'Agata, G. and Ferrara, S., 2005. Gauged supergravity algebras from twisted tori compactifications with fluxes. Nuclear Physics B, 717(1-2), pp.223-245.
\item Reid-Edwards, R. and Spanjaard, B., 2008. $\mathcal{N}$ = 4 gauged supergravity from duality-twist compactifications of string theory. Journal of High Energy Physics, 2008(12), pp.052-052.
\item Dall'Agata, G., Prezas, N., Samtleben, H. and Trigiante, M., 2008. Gauged supergravities from twisted doubled tori and non-geometric string backgrounds. Nuclear Physics B, 799(1-2), pp.80-109.
\item del Moral, M., Heras, C., Leon, P., Pena, J. and Restuccia, A., 2020. Fluxes, twisted tori, monodromy and U(1) supermembranes. Journal of High Energy Physics, 2020(9).
\item Reid-Edwards, R., 2009. Flux compactifications, twisted tori and doubled geometry. Journal of High Energy Physics, 2009(06), pp.085-085.
\item Graña, M., 2006. Flux compactifications in string theory: A comprehensive review. Physics Reports, 423(3), pp.91-158.
\item S. Chiossi and S. Salamon, 2002. The intrinsic torsion of $\text{SU}(3)$ and $G_2$ structures Differential Geometry, Valencia 2001, World Sci. Publishing, pp 115-133.
\item Grana, M., Minasian, R., Petrinin, M. and Tomasiello, A., 2005. Supersymmetric backgrounds from generalized Calabi-Yau manifolds. Fortschritte der Physik, 53(7-8), pp.885-893.
\item F. Witt, Special metric structures and closed forms, Oxford University DPhil thesis (2002) [arXiv:math.DG/0502443].
\item C. Jeschek and F. Witt, 2005. “Generalised G(2)-structures and type IIB superstrings,” JHEP 0503, 053.
\item Graña, M., Minasian, R., Petrini, M. and Tomasiello, A., 2005. Generalized structures of Script N = 1 vacua. Journal of High Energy Physics, 2005(11), pp.020-020.
\item Becker, K. and Tseng, L., 2006. Heterotic flux compactifications and their moduli. Nuclear Physics B, 741(1-2), pp.162-179.
\item Gauntlett, J. P., Martelli, D. and Waldram, D., 2004. Superstrings with intrinsic torsion, Phys. Rev. D 69, 086002.
\item Ivanov, S. and Papadopoulos, G., 2001. Vanishing theorems and string backgrounds, Class. Quant. Grav. 18, 1089. 
\item Blaszczyk, M., Nibbelink, S., Loukas, O. and Ramos-Sánchez, S., 2014. Non-supersymmetric heterotic model building. Journal of High Energy Physics, 2014(10).
\item Nibbelink, S., 2015. Model building with the non-supersymmetric heterotic SO(16)×SO(16) string. Journal of Physics: Conference Series, 631, p.012077.
\item Vafa, C. 2005 The String landscape and the swampland, hep-th/0509212.
\item Palti, E., 2019. The Swampland: Introduction and Review. Fortschritte der Physik, 67(6), p.1900037.

\item Banks, T. and Seiberg, N. 2011 Symmetries and Strings in Field Theory and Gravity, Phys. Rev. D 83 (2011) 084019. 
\item Gaiotto, D., Kapustin, A., Seiberg, N. and Willett, B. Generalized Global Symmetries, JHEP 02 (2015) 172.
\item Banks, T. and Dixon, L. J., Constraints on String Vacua with Space-Time Supersymmetry, Nucl. Phys. B307 (1988) 93–108. 
\item Harlow, D. and Ooguri, H. Symmetries in quantum field theory and quantum gravity, arXiv:1810.05338.

\item Susskind, L. 1967 Trouble for remnants, hep-th/9501106.
\item Israel, W. Event horizons in static vacuum space-times, Phys. Rev. 164 (1967) 1776–1779.
\item Polchinski, J. String theory. Vol. 2: Superstring theory and beyond. Cambridge Monographs on Mathematical Physics. Cambridge University Press, 12, 2007, 10.1017/CBO9780511618123.
\item McNamara, J. and Vafa, C. Cobordism Classes and the Swampland, 1909.10355.
\item Graña, M. and Herráez, A., 2021. The Swampland Conjectures: A Bridge from Quantum Gravity to Particle Physics. Universe, 7(8), p.273.
\item Polchinski, J., 2004. Monopoles, Duality, and String Theory. International Journal of Modern Physics A, 19(supp01), pp.145-154.
\item Arkani-Hamed, N., Motl, L., Nicolis, A. and Vafa, C. 2007 The String landscape, black holes and gravity as the weakest force, JHEP 06, 060.

\item Heidenreich, B., Reece, M. and Rudelius, T. 2017 Evidence for a sublattice weak gravity conjecture, JHEP 08, 025.

\item Aldazabal, G. and Ibanez, L. E., 2018. A Note on 4D Heterotic String Vacua, FI-terms and the Swampland, Phys. Lett. B782, 375–379. 
\item Lee, S. J., Lerche, W. and Weigand, T. 2018. Tensionless Strings and the Weak Gravity Conjecture, JHEP 10, 164.
\item Ooguri, H. and Vafa, C. 2007. On the Geometry of the String Landscape and the Swampland, Nucl. Phys. B 766, 21–33,

\item Lüst, D., Palti, E. and Vafa, C., 2019. AdS and the Swampland. Physics Letters B, 797, p.134867.
\item Ooguri, H. and Vafa, C. 2017. Non-supersymmetric AdS and the Swampland, Adv. Theor. Math. Phys. 21 1787–1801,

\item B. Freivogel and M. Kleban, Vacua Morghulis, 1610.04564.

\item Witten, E. 1982. Instability of the Kaluza-Klein vacuum, Nuclear Physics B 195 481– 492. 

\item Garcia Etxebarria, I. N., Montero, Sousa, M. K. and Valenzuela, I., Nothing is certain in string compactifications, 2005.06494.

\item Guarino, A., Malek, E. and Samtleben, H., 2021. Stable Nonsupersymmetric Anti–de Sitter Vacua of Massive IIA Supergravity. Physical Review Letters, 126(6).

\item Bachas, C. and Lavdas, I., 2018. Massive Anti-de Sitter gravity from string theory. Journal of High Energy Physics, 2018(11).

\item G. Obied, H. Ooguri, L. Spodyneiko and C. Vafa, De Sitter Space and the Swampland, 1806.08362.

\item Ooguri, H., Palti, E., Shiu, G. and Vafa, C. 2019. Distance and de Sitter Conjectures on the Swampland, Phys. Lett. B 788 (2019) 180–184. 
\item Garg S. K. and Krishnan, C. 2019. Bounds on Slow Roll and the de Sitter Swampland, JHEP 11 075.

\item D. Andriot, On the de Sitter swampland criterion, Phys. Lett. B785 (2018) 570–573, [arXiv:1806.10999]. 

\item  Ben-Dayan, I. Draining the Swampland, arXiv:1808.01615. 
\item  Dvali, G., Gomez, C. and Zell, S. 2019. Quantum Breaking Bound on de Sitter and Swampland, Fortsch. Phys. 67 , no. 1-2 1800094. 
\item  Garg, S. K., Krishnan, C. and Zaid Zaz, M. 2019. Bounds on Slow Roll at the Boundary of the Landscape, JHEP 03 029. 
\item  Andriot, D. and Roupec, C. 2019. Further refining the de Sitter swampland conjecture, Fortsch. Phys. 67, no. 1-2 1800105.

\item Quevedo, F., 2002. Lectures on string/brane cosmology. Classical and Quantum Gravity, 19(22), pp.5721-5779.

\item Kachru, S., Kallosh, R., Linde, A. and Trivedi, S., 2003. de Sitter vacua in string theory. Physical Review D, 68(4).
\item Kachru, S., Kim, M., McAllister, L. and Zimet, M., 2021. de Sitter vacua from ten dimensions. Journal of High Energy Physics, 2021(12).
\item Crinò, C., Quevedo, F. and Valandro, R., 2021. On de Sitter string vacua from anti-d3-branes in the large volume scenario. Journal of High Energy Physics, 2021(3).
\item Burgess, C. B. and Quevedo, F., 2022. RG-Induced Modulus Stabilization: Perturbative de Sitter Vacua and Improved $\hbox{D3}$-$\overline{\hbox{D3}}$ Inflation [arXiv:2202.05344 [hep-th]].
\item Landgren, F., 2022. An effective three-dimensional de Sitter cosmology in string theory. Journal of High Energy Physics, 2022(1).
\item Linde, A., 1982. A new inflationary universe scenario: A possible solution of the horizon, flatness, homogeneity, isotropy and primordial monopole problems. Physics Letters B, 108(6), pp.389-393.
\item Albrecht, A. and Steinhardt, P., 1982. Cosmology for Grand Unified Theories with Radiatively Induced Symmetry Breaking. Physical Review Letters, 48(17), pp.1220-1223.
\item Das, S., 2020. Distance, de Sitter and Trans-Planckian Censorship conjectures: the status quo of Warm Inflation, Phys. Dark Univ. 27 100432
\item Berera, A. and Calderon, J. R., 2019. Trans-Planckian censorship and other swampland bothers addressed in warm inflation, Phys. Rev. D 100 123530
\item Brandenberger, R., Kamali, V. and Ramos, R., 2020. Strengthening the de Sitter swampland conjecture in warm inflation. Journal of High Energy Physics, 2020(8).
\item Rasulian, I., Torabian, M. and Velasco-Sevilla, L., 2021. Swampland de Sitter conjectures in no-scale supergravity models. Physical Review D, 104(4).
\item Wetterich,C., 1998 , Cosmology and the fate of dilatation symmetry, Nuclear Physics B 302, no. 4 668– 696.
\item  Ratra, B. and Peebles, P. J. E., 1998. Cosmological consequences of a rolling homogeneous scalar field, Phys. Rev. D 37 3406–3427. 
\item Copeland, E. J., Sami, M., and Tsujikawa, S., 2006. Dynamics of dark energy, Int. J. Mod. Phys. D15 (2006) 1753–1936.
\item Damian, C. and Loaiza-Brito, O., 2018. Two-Field Axion Inflation and the Swampland Constraint in the Flux-Scaling Scenario. Fortschritte der Physik, 67(1-2), p.1800072.
\item Achcarro, A. and Palma, G. A., 2019. The string swampland constraints require multi-field inflation, JCAP 1902 041.
\item Achucarro, A., Copeland, E. J., Iarygina, O., Palma, G. A., Wang, D. G. and Welling, Y., 2020. Shift-Symmetric Orbital Inflation: single field or multi-field?, Phys. Rev. D 102, 021302.
\item Matsui, H. and Takahashi, F., 2019. Eternal Inflation and Swampland Conjectures, Phys. Rev. D99, no. 2 023533,
\item Kinney, W. H., 2019. Eternal Inflation and the Refined Swampland Conjecture, Phys. Rev. Lett. 122, no. 8 081302
\item Heisenberg, L., Bartelmann, M., Brandenberger, R. and Refregier, A., 2019. Horndeski gravity in the swampland. Physical Review D, 99(12).
\item Artymowski, M. and Ben-Dayan, I., 2019. f(R) and Brans-Dicke Theories and the Swampland, JCAP05(2019)042.
\item Brahma, S. and Hossain, M., 2019. Dark energy beyond quintessence: constraints from the swampland. Journal of High Energy Physics, 2019(6).
\item Laliberte, S. and Brandenberger, R., 2020. String gases and the swampland. Journal of Cosmology and Astroparticle Physics, 2020(07), pp.046-046.
\item Bidussi, L., Harmark, T., Hartong, J., Obers, N. and Oling, G., 2022. Torsional string Newton-Cartan geometry for non-relativistic strings. Journal of High Energy Physics, 2022(2).
\item Oling, G. and Yan, Z., 2022. Aspects of Non relativistic strings.  https://doi.org/ 10.3389/fphy.2022.832271.
\item Gomis, J. and Ooguri, H. (2001). Nonrelativistic closed string theory, Journal of Mathematical Physics, 42(7), pp. 3127–3151.  https: //doi.org/10.1063/1.1372697.
\item Gomis, J., Yan, Z. ; Yu, M., 2021. Nonrelativistic open string and Yang-Mills theory. Journal of High Energy Physics, 2021(3). 

\item Gomis, J., Oh, J. ; Yan, Z., 2019. Nonrelativistic string theory in background fields. Journal of High Energy Physics, 2019(10).

\item Bergshoeff, E.A. et al., 2019. String theory and string Newton–cartan geometry. Journal of Physics A: Mathematical and Theoretical, 53(1), p.014001. 

\item Kluson, J., 2019. Nonrelativistic string theory sigma model and its canonical formulation. The European Physical Journal C, 79(2).

\item Kluson, J., 2021. D-brane actions in nonrelativistic string theory and T-duality. Physical Review D, 104(8). 

\item Gomis, J., Yan, Z. ; Yu, M., 2021. T-duality in nonrelativistic open string theory. Journal of High Energy Physics, 2021(2). 
\item Bergshoeff, E., Gomis, J. and Yan, Z., 2018. Nonrelativistic string theory and T-duality. Journal of High Energy Physics, 2018(11).
\item AbdusSalam, S., Abel, S., Cicoli, M., Quevedo, F. and Shukla, P., 2020. A systematic approach to Kähler moduli stabilisation. Journal of High Energy Physics, 2020(8).
\item Cicoli, M., Schachner, A. and Shukla, P., 2022. Systematics of type IIB moduli stabilisation with odd axions. Journal of High Energy Physics, 2022(4).
\item Grimm, T., Plauschinn, E. and van de Heisteeg, D., 2022. Moduli stabilization in asymptotic flux compactifications. Journal of High Energy Physics, 2022(3).
\item Andriot, D., 2019. Open problems on classical de Sitter solutions, Fortsch. Phys. 67 1900026.

\end{enumerate}

\end{document}